\documentclass[pdflatex,sn-mathphys-num]{sn-jnl}% Math and Physical Sciences Numbered Reference Style
\usepackage{graphicx}%
\usepackage{multirow}%
\usepackage{amsmath,amssymb,amsfonts}%
\usepackage{amsthm}%
\usepackage{mathrsfs}%
\usepackage[title]{appendix}%
\usepackage{xcolor}%
\usepackage{textcomp}%
\usepackage{manyfoot}%
\usepackage{booktabs}%
\usepackage{algorithm}%
\usepackage{algorithmicx}%
\usepackage{algpseudocode}%
\usepackage{listings}%

\newcommand{\avg}[1]{\left< #1 \right>}
\newcommand{\avgE}[1]{\mathbb{E}\left({#1}\right)}

\theoremstyle{thmstyleone}%
\theoremstyle{thmstyletwo}%

\theoremstyle{thmstylethree}%

\begin{document}

\title[Debt, Growth, and the Carbon Lock-In]{Debt, Growth, and the Carbon Lock-In}

\author[a, b, 1]{Silvia Montagnani}\equalcont{These authors contributed equally to this work.}
\author*[c, 1]{Barnabe Ledoux}\email{barnabeledoux@gmail.com}\equalcont{These authors contributed equally to this work.}
\author[c]{David Lacoste}

\affil[a]{iRisk, LEM (CNRS UMR 9221) / I\'ESEG School of Management, Universit\'e de Lille, 59000 Lille, France}
\affil[b]{Center for Critical Computational Studies (C\textsuperscript{3}S), Goethe University Frankfurt, 60323 Frankfurt, Germany}
\affil[c]{Gulliver Laboratory, UMR CNRS 7083, PSL Research University, ESPCI, 75231 Paris, France}

% Please give the surname of the lead author for the running footer

 \abstract{Despite decades of climate policy and rapid improvements in energy efficiency, global CO$_2$ emissions continue to rise, suggesting the presence of structural drivers that offset efficiency gains. Here we identify financial leverage as a key mechanism underpinning this persistent overshoot.

We develop a stochastic macro-financial model linking credit dynamics, economic growth, bankruptcy risk, and cumulative carbon emissions. The model shows that debt-financed growth systematically amplifies cumulative emissions, locking economies into high-carbon trajectories even as emissions intensity declines. This arises from a double constraint: debt repayment requires sustained growth, while growth remains energy-dependent and thus generates emissions. When growth becomes increasingly dependent on leverage, financial instability and cumulative emissions rise, while gains in real wealth diminish, revealing a leverage frontier beyond which additional credit primarily generates risk.

Calibrating the model to multi-decade data for the United States, China, France and Denmark, we find a robust coupling between debt accumulation, cumulative GDP and cumulative emissions across distinct economic structures. These results challenge the feasibility of growth–emissions decoupling under prevailing credit-driven growth regimes and indicate that achieving net-zero targets requires aligning credit allocation with decarbonisation objectives.}


\maketitle

\subsection*{Introduction}
Debt is a social construct~\cite{graeber_debt_2011, hagens_economics_2020}, and may thus appear disconnected from the physical reality. However, debt has real-world implications because borrowed money facilitates the growth of businesses including military activities by enabling the consumption of energy and resources. At the same time, it can also cause financial instabilities due to leverage cycles 
~\cite{keen_finance_1995, keen_predicting_2013, schularick_credit_2012, keen_emergent_2020, giraud_macrodynamics_2023}.
Thus, debt brings consumption and production forward by enabling energy investment in resource-intensive sectors, including marginal resource extraction and high-cost infrastructure~\cite{hagens_economics_2020}.

\begin{figure}[ht]
    \centering
    \includegraphics[width=0.67\textwidth]{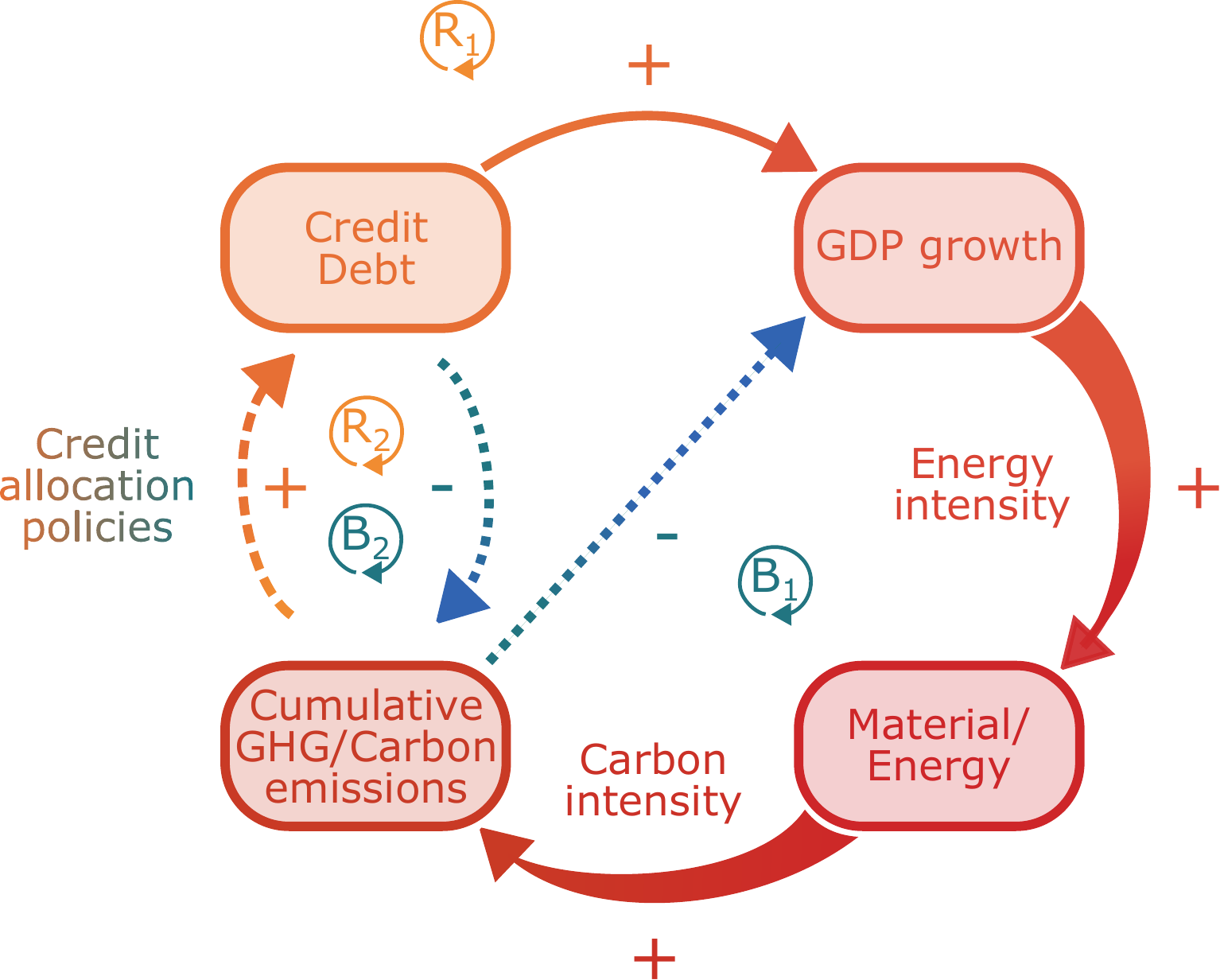}
        \caption{\textbf{The debt-carbon spiral: leverage enhances economic growth, but it also increases emissions and default risk, thereby establishing a feedback loop that locks in carbon-intensive development and jeopardizes climate targets and financial stability.}
        In this scheme, the letters (R) and (B) represent reinforcing and balancing loops, respectively. The primary reinforcing loop (R1) illustrates how the expansion of credit and leverage stimulates GDP growth, leading to increased energy consumption and CO$_2$ emissions. A secondary reinforcing loop (R2) describe the necessity of additional borrowing to maintain the same level of resource extraction as the energy return on investment (EROI) declines and the costs relative to climate change damages rise~\cite{hall_peak_2018, kotz_economic_2024}. This, in turn, leads to more aggressive fiscal and credit policies that further intensify debt-financed growth. Two balancing loops counteract these self-propelling cycles: (B1) captures the negative economic effects of environmental externalities, such as pollution, climate-related losses, and reduced productivity, that constrain growth~\cite{kotz_economic_2024}, while (B2) captures technological and structural decarbonization mechanisms that reduce the carbon intensity of energy use and improve efficiency. Feedback that are qualitatively supported in the literature but are not explicitly quantified in our model are indicated by dashed lines.}
    \label{fig:fig1}
\end{figure}

In fact, debt creates new feedback loops in an economy as illustrated  in Fig.~\ref{fig:fig1}. In such a loop, credit expansion stimulates growth, thereby increasing  energy demand~\cite{hall_peak_2018, kotz_economic_2024} (often directed towards carbon-intensive sectors) and results in carbon emissions, despite ongoing efforts towards decarbonization. 
In the literature, this feedback loop is known as the 'macroeconomic rebound effect' \cite{dafermos_stock-flow-fund_2017}, when investments in clean energy boost productivity and growth, but also increase overall emissions as a result.

Similar feedback loops exist in other fields such as biology.
In biology, living cells adjust to environmental constraints by employing two complementary strategies: (i) decreasing their metabolic rate to consume less energy per unit time, or (ii) reconfiguring their metabolic pathways to optimize the use of available energy or alternative sources~\cite{brown_energetic_2011, hagman_study_2015, woronoff_metabolic_2020, biselli_slower_2020}.
In a similar manner, economies that are confronted with resource constraints have the option to either (i) decrease the energy required per unit of output, thereby reducing the carbon intensity of GDP, or (ii) transition their energy mix to low-carbon sources~\cite{loreau_biodiversity_2021}.
An important limitation to this analogy between biology and economy, however, is that there is no biological equivalent of debt. Debt is thus inherent to financial systems; it facilitates growth on the promise of future expected gains in a manner that biological systems cannot, thereby establishing new mechanisms of instability due to an increase of emissions, resource depletion, socio-economic destabilization and
and ecosystem degradation~\cite{marsden_financial_2024, quilcaille_systematic_2025}. 
 
%The declining carbon intensity, $I(\tau)$, can represent both strategies in our model. 
%Ultimately, our research demonstrates that technological advancements alone are insufficient to mitigate the emissions associated with economic activity when leverage is too large.
%Our interest in the topic was sparked by the fact that, unlike the metabolism of living organisms, there is no biological equivalent of debt.

Motivated by this observation, we study in this paper the link between finance and climate change, focusing particularly on the role of debt. 
Another motivation for this study is that standard Integrated Assessment Models (IAMs), which typically represent emissions as a function of exogenous GDP growth ~\cite{raupach_global_2007, nordhaus_climate_2017}, do not model the financial system nor investors’ decisions. Thus, the
feedback loop represented in Fig.~\ref{fig:fig1} that couples the financial system and mitigation pathways is typically not taken into account by IAMs. Several authors have identified this shortcoming and have criticized IAMs for their lack of a financial system \cite{giraud_macrodynamics_2023,battiston_accounting_2021}.

To capture these feedback effects, we build a stochastic credit-risk model by combining deterministic macroeconomic models \cite{dallery_comptabilite_2023, godley_monetary_2007, keen_predicting_2013,keen_emergent_2020}, with ingredients from Kelly's gambling model~\cite{kelly_jr_new_1956}.
%, because this model offers a general theory of money, which is not based on notions of economic equilibrium. 
In 1956, Kelly found a deep connection between information theory and gambling in horse races. This work then became highly influential in investment science~\cite{maclean_kelly_2011, kim_kelly_2024} and in evolutionary biology~\cite{donaldson-matasci_phenotypic_2008}. 
%Kelly's optimal strategy defines an optimal capital allocation by maximizing the expected logarithmic growth rate.
The model demonstrates that diversification of assets enables a mitigation of risk from environmental fluctuations, a goal that is also central in biology and ecology and is known under the name of bet-hedging~\cite{loreau_biodiversity_2021, donaldson-matasci_phenotypic_2008, de_groot_effective_2023}. 
In this analogy, the growth of the gambler's capital is equivalent to the growth of the population, and the risk of bankruptcy is comparable to the extinction of the population. Kelly's optimal strategy is known to be risky, and this observation hints at a general trade-off, the risk-return trade-off, between the risk of bankruptcy and the average long-term growth rate of the capital~\cite{dinis_phase_2020, cavallero_trade-off_2025}. 

%In finance, investors borrow money to invest, and as a result, their capital can grow or decline if they fail to repay their debts. 
The original Kelly's model, however, does not include the possibility of bankruptcy nor the possibility of borrowing (debt). To address this issue, we build a stochastic credit-risk model, that merges the stochastic nature of growth in  Kelly's model with deterministic credit-risk models \cite{keen_predicting_2013,keen_emergent_2020}.
%To model these effects, we extend Kelly’s framework to include default dynamics under compounding interest obligations. 
In this way, once the borrower has taken out a loan, the debt accumulates with an interest rate at each time step, while the capital evolves stochastically. When the better pays back the debt, the capital is reset at a lower value. The evolution of the capital can thus be compared to a biased random walk with resetting~\cite{majumdar_universal_2010, majumdar_random_2015, flaquer-galmes_intermittent_2024}, or to a coupled dynamics of money and anti-money variables~\cite{schmitt_statistical_2014}.
%In this paper, we study the optimal payback time depending on the parameters of the model. This question is somewhat related to the issue of finding the optimal time to buy or sell a stock in the Black-Scholes model~\cite{majumdar_optimal_2008}.

In the following, we use this stochastic credit-risk  model to investigate the role of credit expansion in driving economic growth and cumulative CO$_2$ emissions.
Besides quantifying this causal relationship, our stochastic framework also establishes an optimal leverage frontier: the probability of long-term solvency approaches zero when the intrinsic growth rate (growth that originates within an economy, fueled by factors such as productivity improvements) falls below the effective interest rate, suggesting that structural financial instability is the ultimate effect of excessive borrowing.
We also show that credit is the primary driver of several major economies and that this comes at the cost of increasing global dissipation in the form of emissions
%. This framework thus exhibits double materiality: financial leverage is responsible for driving emissions (inside–out impact), while climate instability and policy responses amplify financial risk (outside–in impact)
~\cite{palma_path_2025, battiston_accounting_2021}. We used country-level economic data to predict the cumulative emissions per capita for these countries. Cumulative CO$_2$ emissions, rather than CO$_2$ emissions, are a key quantity that captures the effect of economic activities in increasing global warming. We demonstrate that to fuel economic growth, credit is required, which has a cost in terms of CO$_2$ emissions. %Debt is therefore a means of consuming more at a given time than the economy can sustain without credit~\cite{hagens_economics_2020}, which leads to the depletion of resources sooner, typically with a lower efficiency (or with a higher carbon intensity).  

Our contribution complements standard Integrated Assessment Models (IAMs), which typically do not aim to reproduce past emissions, GDP co-evolution, or assess the causal role of credit in driving the rise in emissions. 
    In contrast, our model uses empirical fiscal, macroeconomic, and climate data to explicitly link debt-financed growth with cumulative emissions, thereby offering a plausible explanation for the persistent proportionality between cumulative GDP and emissions, thus assessing the responsibility for cumulative carbon emissions even in countries with fast decarbonization trajectories. Our approach aligns more closely with recent works in macro-finance that analyze climate-related financial risk~\cite{battiston_accounting_2021}, and with studies in ecological macroeconomics \cite{rezai_ecological_2013, dalessandro_feasible_2020} and system dynamics \cite{lenton_spread_2024} that attempt to take into account socio-environmental feedback loops similar to the debt-carbon spiral presented in this paper (Fig.~\ref{fig:fig1}).

\subsection*{Model}

We consider an investor who can borrow money at each time step $\tau$ with the interest rate $\rho_{\tau}-1$ (real interest rate accounting for inflation \cite{azoulay_chapitre_2023}), thus increasing both its capital $C_{\tau}$ and its debt $D_{\tau}$. The amount of borrowed money is quantified by the leverage $L_{\tau}$, which is defined as the ratio of the asset value to the cash needed to purchase it \cite{schularick_credit_2012}. Since we consider an isolated investor, borrowing to an external actor, the leverage can be larger than $1$, whereas in a closed economic system, the total leverage should be $1$. The evolution of the capital is stochastic as in Kelly's model~\cite{maclean_kelly_2011, kim_kelly_2024}, and is quantified by the intrinsic growth rate (or capital return) $\gamma_{\tau} - 1$. 
The model does not explicitly describe variables such as labor, employment rate or wages, all of which can exhibit complex dynamics \cite{keen_finance_1995, keen_predicting_2013, giraud_macrodynamics_2023}. Instead, it focuses on the coupling between the capital and the debt. 
At each time step, the investor can pay back their debt, and
bankruptcy occurs if the debt exceeds the capital when repayment is due.
%Without debt, the gross capital cannot hit $0$ in a multiplicative model (such as Kelly's model), and bankruptcy never occurs. 
%In this model, the capital can become negative if the debt exceeds the capital at the time the debt is repaid. 
Therefore \textit{solvency} means that the debt can always be paid back because ($\gamma_{\tau}C_{\tau} > D_{\tau+1}$). We call $\gamma_{\tau}C_{\tau}-D_{\tau+1}$ the \textit{net capital}, whose sign indicates if bankruptcy will happen in case of payback. In the following sections, we outline two different strategies that represent distinct real-world debt-financed growth regimes. 
The mathematical model is further detailed in Methods and in  sections 1.4 and 1.5 of SI where we explain how our framework relate to known macro-economical models~\cite{keen_finance_1995, keen_emergent_2020, keen_predicting_2013, dallery_comptabilite_2023, godley_monetary_2007}. 

\subsubsection*{Finite leverage --- one time borrowing (strategy A)}
First, we assume that the investor initially borrows money but never borrows afterwards. We refer to this strategy as strategy A, which is akin to an individual taking out a loan to buy a house. In this case, the gross capital evolves as in the classical Kelly's model, and the debt evolves with a constant interest rate $\rho-1$. At the time of payback, the capital can become negative, and in that case, bankruptcy occurs. Under the assumption of no bankruptcy, we can define the long-term growth rate as the growth rate of the capital when $\tau \to \infty$. However, there is a non-zero probability of hitting bankruptcy when the debt is paid back, and the larger the leverage, the more likely the bankruptcy in this model. 

\subsubsection*{Persistent leverage --- periodic borrowing (strategy B)}

Let us instead assume that the investor borrows money at each time step, with leverage $L_{\tau}$ (now time-dependent), which we refer to as strategy B. This strategy is rather that of a state or a company taking on credit yearly to ensure its budget balance. In this case, before payback, the gross capital evolves as in the standard Kelly's model, with a modified stochastic growth rate, and the debt still evolves with an interest rate $\rho-1$. We can separate the value added from production, and that which is due to borrowing in the dynamics,  and define the stochastic \textit{intrinsic growth rate} as the growth rate to which the contribution of  borrowing~\cite{dallery_comptabilite_2023} is subtracted.

\section*{Results}

\subsection*{Lock-in of carbon emissions}

Now, we scale up the above model based on strategy B to describe the wealth and the carbon emissions of a given country. Let $C_{\tau}$  represent the economic wealth in the year $\tau$ of that country, and let us assume that an external bank lends money each year to that country. As a result, its debt, encapsulated by the variable $D_{\tau}$, increases. Let us also assume that the leverage is constant within the time frame considered. Empirically, the estimated leverage $L_{\tau}$ remains indeed fairly stable within each economy, varying by only a few percent across years (as shown in Fig.~S1 in the Supplementary Information). This supports the hypothesis of a quasi-stationary financial amplification of growth.

In practice, the GDP during a given period is often used as a measure of economic wealth and has also been criticized for this reason~\cite{giannetti_review_2015, lianos_adjusting_2021}. %It consists of summing the monetary values of all products created; therefore, the added growth product between time $\tau$ and $\tau+1$ is $C_{\tau+1}-C_{\tau}$. 
With that in mind, we further assume that this economic wealth represents the capital invested in various industries that consume energy to produce carbon emissions at time $\tau$, described by the function $\epsilon(\tau, C_{\tau})$. This function should be an increasing function of the capital, as observed in practice~\cite{tucker_carbon_1995}, and its average should increase with leverage (i.e., the ratio of total money invested to money borrowed). 
%We ultimately have that $\epsilon$ is an increasing function of the capital $C_{\tau}$, which is observed in practice~\cite{tucker_carbon_1995}.

We connect the emissions to the carbon intensity of the economy thanks to the Kaya relationship~\cite{friedlingstein_persistent_2014, tavakoli_journey_2018,raupach_global_2007, abbasi_carbon_2022}:
\begin{equation}
    \epsilon (\tau, C_{\tau}) = C_{\tau} \times I(\tau),
\end{equation}
where  $I(\tau)$ is the carbon intensity of the economy, measured by  the carbon cost of one unit of $GDP$ \cite{friedlingstein_persistent_2014}. 
The parameter $I(\tau)$ is typically a decreasing function of $\tau$~\cite{friedlingstein_persistent_2014,raupach_global_2007,abbasi_carbon_2022} because energy mixes tend to incorporate more renewable energies over time. Note that $I(\tau)$ can decrease for two main reasons, as illustrated in Fig.~\ref{fig:fig1}, either the energy cost of $1 USD$ decreases, meaning that money is used in less energy-intensive activities, or the carbon cost of energy decreases, meaning that the energy mix of a country includes "greener" energies. However, this does not mean that cumulative emissions are reduced, as we will see. 
%This is analogous to a living cell, which can cope with an energy poor environment, either by slowing down growth (adjusting the amount of energy used per unit time) or by altering its metabolic pathway (changing the way energy is allocated).

%This function represents how GDP is allocated and what are the resulting emissions. 
Let us now examine the lock-in effect arising from path-dependent cumulative emissions. Public debt can be a powerful tool for investing in the transition, but it can also be used to pursue carbon-intensive economic activities; the difference between the two is encapsulated in the concept of carbon intensity. As this quantity changes with time~\cite{friedlingstein_persistent_2014}, the cumulative emissions corresponding to a particular trajectory for a country's GDP will also depend on the trajectory of the carbon intensity of this economy. We can measure the lock-in effect using the \textit{path-dependent intensity}:

\begin{equation}
    \mathfrak{I}_{\tau}\left(\{C_{\tau}\}\right) = \frac{\sum_{\tau'=0}^{\tau} I(\tau')C_{\tau'}}{\sum_{\tau'=0}^{\tau}C_{\tau'}},
\end{equation}
which compares cumulative carbon emissions to cumulative GDP over a given period. This quantity represents the cumulative carbon emissions for a given amount of money invested. Contrary to carbon intensities, this variable is path-dependent, similarly to heat and work in thermodynamics~\cite{seifert_stochastic_2025}. 
%This quantity is accurate to study efforts done to decarbonize the economy by tracking the overall effect of energy transition in a country over a period. 
This quantity allows for assessing the actual responsibility of countries for cumulative carbon emissions over a given period, irrespective of recent decarbonization efforts captured by carbon intensity.

In SI 2.3, we demonstrate that if carbon intensities $I(\tau)$ strictly decrease, then  $\mathfrak{I}_{\tau}\left(\{C_{\tau}\}\right)$ also  decreases over time. We then study the influence of the decarbonization rate (or efficiency of innovations) in SI 2.3 and conclude that if $I(\tau)$ is decreasing linearly, the difference between $\mathfrak{I}_{\tau}\left(\{C_{\tau}\}\right)$ and $I(\tau)$ stabilizes to a finite value. 
%In general, the difference goes like $-(dI/d\tau)/(e^{r}-1)$ at large times. 
If $I(\tau)$ decreases faster than linearly, the difference tends to $+\infty$, whereas if it decreases more slowly than linearly, the difference first increases and then tends to $0$. In practice, this means that the faster decarbonization occurs, the more early carbon-intensive years will matter in the path-dependent intensity. On the contrary, for slow decarbonization, recent years matter more in the path-dependent intensity, making it closer to the carbon intensity.

%As $\mathfrak{I}_{\tau}\left(\{C_{\tau}\}\right)$ depends on two integrated quantities, it reflects the history of an economy's evolution. 

\begin{figure*}
    \centering
    \includegraphics[width= \linewidth]{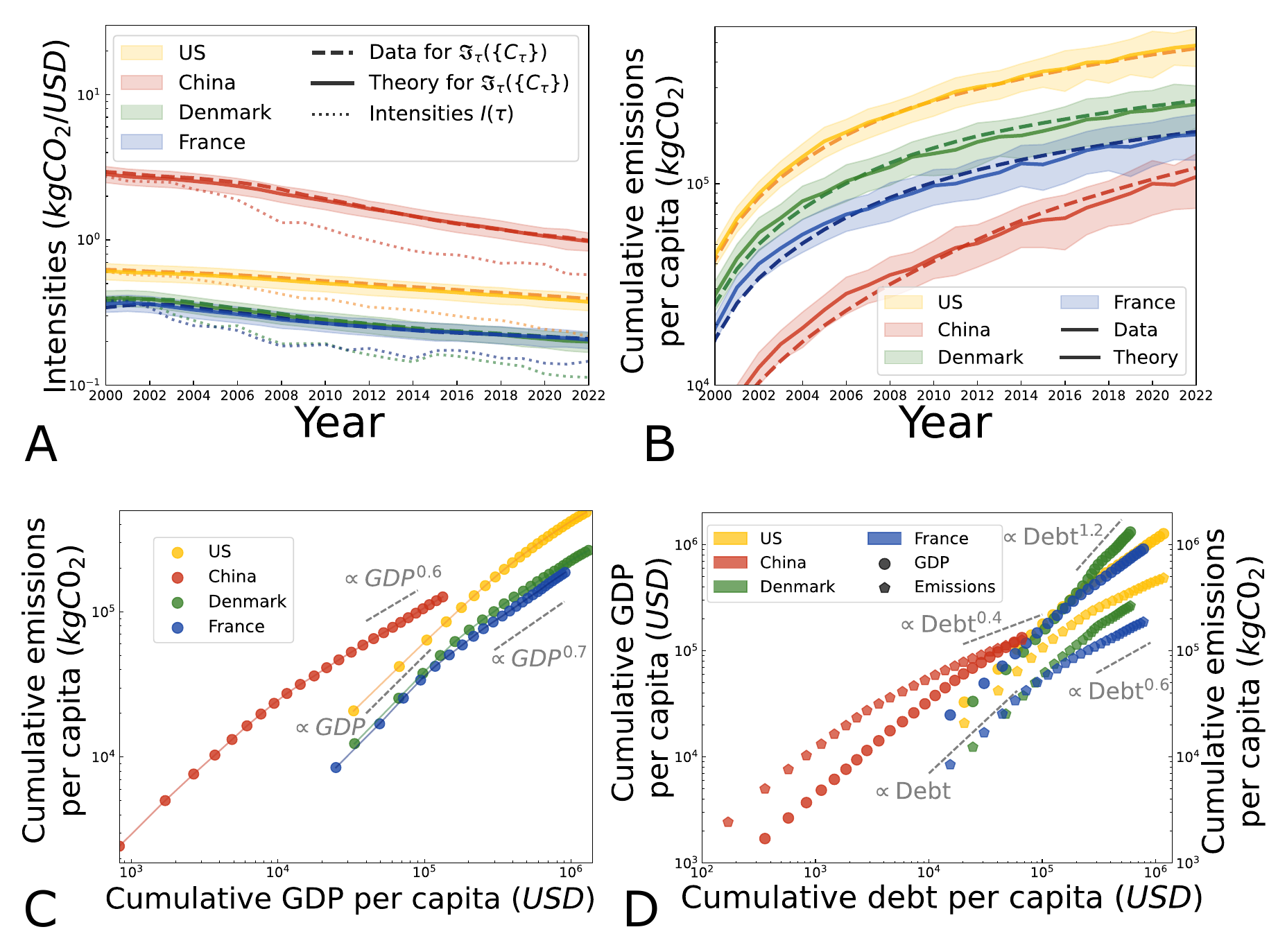}
    \caption{\textbf{A} Path-dependent intensities $\mathfrak{I}_{\tau}\left(\{C_{\tau}\}\right)$ starting from year 2000 and yearly carbon intensities $I(\tau)$ (consumption-based) for different countries. Data for path-dependent intensities and yearly carbon intensities are shown as dotted lines.
    The theoretical prediction   obtained by computing $C_{\tau}$  from our model, and combining it with $I_{\tau}$ from data, is 
shown as the solid line representing the average trend, with the colored area around that line representing the standard deviation.
%Even if yearly carbon intensities tend to decrease (the carbon cost of $1 USD$ invested decreases with time), path-dependent intensities decrease more slowly than intensities. %Therefore, path-dependent intensities can almost be considered as constant on a yearly timescale. 
In practice, we used country-specific measured values of the \textit{intrinsic growth rate}, interest rate, and leverage, to estimate $C_{\tau}$. \textbf{B} Comparison of cumulative $CO_2$ emissions (data in dotted lines) for different countries (US, China, Denmark, and France) to the predictions of our stochastic model as solid lines, with standard deviation represented by the colored area. We report the mean average errors (MAE) and coefficients of determination between the theory and data for plot A and B in Supplementary Information. \textbf{C} Relationship between cumulative emissions per capita and cumulative GDP per capita from the data: we observe a clear increasing trend between both, with a slope that varies according to the country. As in Fig A, one observes that  that China has the most carbon-intensive economy, while the US has the highest cumulative emissions, owing to its high cumulative GDP per capita. \textbf{D} Correlation between cumulative GDP and cumulative debt (left axis), and correlation between cumulative emissions and cumulative debt (right) axis. Cumulative GDP (circles) is mostly a linear function of cumulative debt, except for Denmark, where it is slightly faster than linear. Concerning cumulative emissions (pentagons), the trend is initially linearly correlated with cumulative debt, then becomes slower than linear (in particular for China). Again, this is an effect of the decrease in carbon intensity with time.}
    \label{fig:fig2}
\end{figure*}

A higher path-dependent intensity strengthens the lock-in effect, since $CO_2$ emissions cannot be reset to zero annually.
We show path-dependent intensities, as well as intensities in Fig.~\ref{fig:fig2}. We observe that the orders of magnitude differ widely among countries. We also observe that even if carbon intensities decrease over time, path-dependent intensities remain practically unchanged over a 20-year period. This means that cumulative emissions are largely proportional to cumulative GDP, despite efforts to reduce the annual carbon intensity.

The difference between annual carbon intensity and path-dependent intensity is studied more closely in the Supplementary Information. In Fig.~\ref{fig:fig2}C, we show the correlation between cumulative emissions per capita (since 1998) and cumulative GDP (since 1998) for each country. When the carbon intensity decreases with time while the economy grows, $\mathfrak{I}_{\tau}\left(\{C_{\tau}\}\right)$ is larger than $I(\tau)$. The difference between these two quantities depends on the long-term growth rate $W$ and on the amplitude of the fluctuations in the economy measured by  $\sigma_X$.
We observe that even under steady decarbonization, the ratio of cumulative emissions to cumulative GDP remains higher than the annual carbon intensity. 
Moreover, larger economic fluctuations - as captured by $\sigma_X$ - amplify the difference between $\mathfrak{I}_{\tau}\left(\{C_{\tau}\}\right)$ and $I(\tau)$. 
This phenomenon arises because periods of economic expansion, characterized by high carbon intensity, are followed by recessions with lower carbon intensity.
In this case, reducing the annual carbon intensity has little effect on the path-dependent carbon intensity, yielding a larger degree of carbon lock-in.

The trends for $\mathfrak{I}_{\tau}\left(\{C_{\tau}\}\right)$ and $I(\tau)$ are similar across all countries, with varying carbon intensities. For countries with small GDP per capita, the relationship is almost linear but bends as a power law with an exponent slightly smaller than one at larger GDP per capita, as shown in Fig.~\ref{fig:fig2}C. The lower this exponent, the more significant the efforts to decarbonize the economy.

By shifting consumption forward and increasing annual GDP per capita, debt increases cumulative emissions. This result is robust even if debt is used to decrease the annual carbon intensity, $I(\tau)$, because cumulative emissions are path-dependent.

\subsection*{Energetic and dissipative cost of an economy}
\begin{figure*}
    \centering
    \includegraphics[width=\textwidth]{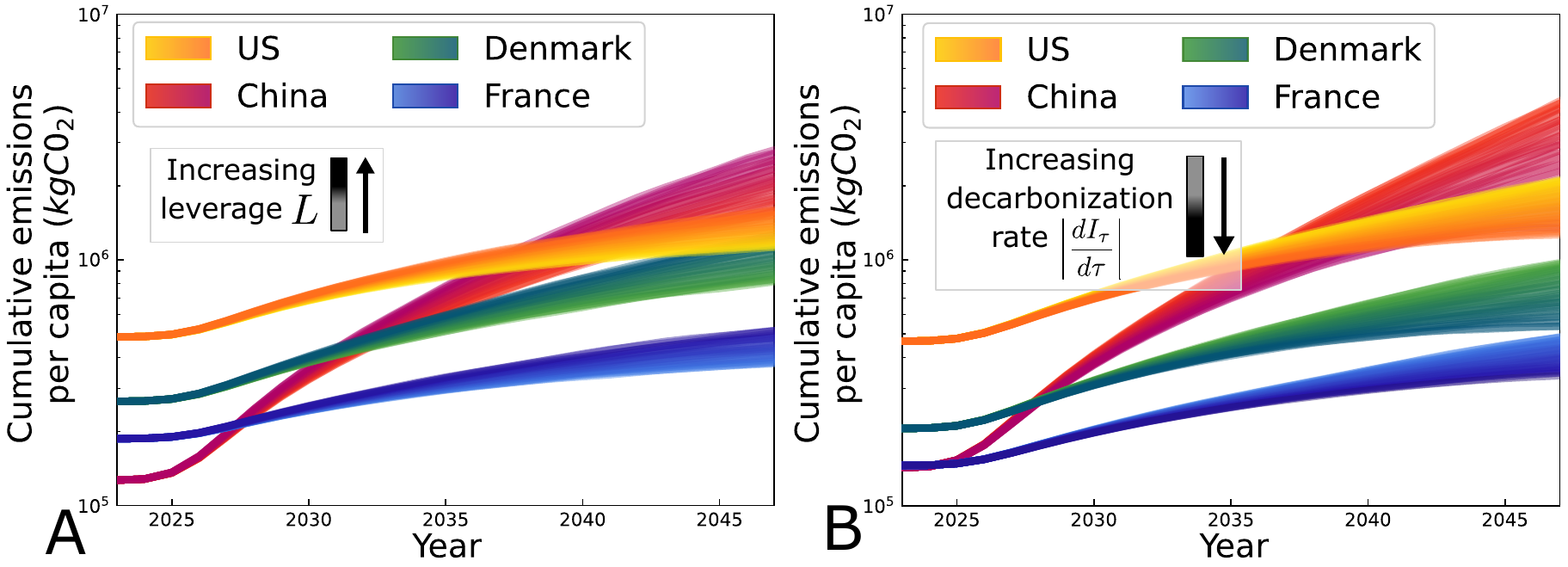}
    \caption{\textbf{A} \textit{Effect of leverage}: predictions for cumulative emissions (starting from 1998) depending on leverage in different countries. For each country, the color gradient indicates increasing leverage from $1.00$ to $1.05$, resulting in higher cumulative emissions per capita. A linearly decreasing carbon intensity has been assumed, with different parameters for each country~\cite{friedlingstein_persistent_2014}, based on consumption-based emissions from 1998 to 2022. We also assume that the distributions of $\gamma$ remain constant for each country. We use $W=\avg{\ln(\gamma_{\tau})}$ and $\sigma_{X}$ (see values in Methods), as well as a fixed value of leverage, as inputs for the model. As expected, the larger the leverage, the greater the potential for GDP growth, leading to higher cumulative emissions despite decreasing carbon intensities. Changes by a few percent in leverage are enough to trigger changes of almost $1$ order of magnitude in cumulative emissions per capita after $25$ years. \textbf{B} \textit{Effect of policies}: Predictions for cumulative emissions (starting from 1998) depending on how strong the effort is to reduce carbon intensity. For each country, the color gradient corresponds to increasing $\left|dI_{\tau}/d\tau\right|$ (where $dI_{\tau}/d\tau$ is negative), indicating an increasing effort to reduce the carbon intensity of the economy. $\left|dI_{\tau}/d\tau\right|$ varies between $0$ and $1.5$ times the current trend for each country. We observe that for high enough values of $\left|dI_{\tau}/d\tau\right|$, $I_{\tau}$ becomes negative (carbon capture), leading to decreasing cumulative emissions. The leverage is taken equal to the average leverage measured over the period $1998-2022$ for each country.}
    \label{fig:fig3}
\end{figure*}

Considering a period $t$, we can write the cumulative emissions $\mathcal{E}(t)$ as:
%and resource consumptions $\mathcal{R}(t)$ as 
\begin{equation}
\begin{split}
    \mathcal{E}(t) &= \sum_{\tau \leq t} \epsilon(C_{\tau},\tau),
 %   \\ \mathcal{R}(t) &= \sum_{\tau \leq t}r(C_{\tau},\tau),
\end{split}
\end{equation}

assuming in a first approximation that yearly carbon intensity $I(\tau)$ does not depend explicitly on $C_{\tau}$, it has historically been measured as almost linear in most countries~\cite{friedlingstein_persistent_2014}. Therefore, writing $I(\tau)=I_0 - \eta (\tau-\tau_0)$,

\begin{equation}
    \mathcal{E}(t) = I_0\sum_{\tau \leq t} C_{\tau} - \eta\sum_{\tau \leq t} \tau C_{\tau}.
\label{eq:emissions}
\end{equation}

Using this formula for the US $GDP$ data between 1990 and 2020, we obtain cumulative carbon emissions of $\mathcal{E} = 7\times10^5 kg \, CO_2/\text{capita}$, close enough to the real value of $6\times10^5 kg \, CO_2/\text{capita}$ on Fig.~\ref{fig:fig2}B.

We have analyzed data for the US deficit~\cite{noauthor_monthly_2025} and found that the added deficit in 2023 is $1.7 \times 10^{12} USD$, resulting in a deficit per capita of $ 5.1 \times 10^3 USD$ in 2023. From this, we can obtain a value of the leverage in 2023 of $L=1.06$. If we use strategy B to describe the growth of the US economy, the \textit{intrinsic growth rate} (that is, the growth rate minus debt-funded spending) is:

\begin{equation}
    \gamma_{eff,\tau} = \frac{C_{\tau+1}-C_{\tau}}{C_{\tau}} - (L-1).
\end{equation}

Studying this quantity allows for the decorrelation of the value added from credit to the added value (accounting for depreciation and production). On average, since 1980, this value is close to 0. Using data~\cite{noauthor_us_2025, noauthor_us_2025-1, noauthor_historical_2025} for US GDP per capita since 1960, we can compute its growth rate. However, as mentioned earlier, we must distinguish the \textit{intrinsic growth rate} from growth due to credits. %To do this, we get the growth rate of the capital by evaluating the evolution after each year, then taking the average of the $\ln$ of the distribution of growth rates, the $\ln$ of the average of the distribution of growth rates, and the standard deviation of the $\ln$ of the distribution. 

While the public deficit can stimulate economic activity within a Keynesian framework~\cite{roge_capitalisme_2023, guy_renouveau_2023} (national income is proportional to government spendings through a Keynesian multiplier \cite{godley_monetary_2007, dalziel_keynesian_1996}), it also increases cumulative carbon emissions in proportion to the path-dependent carbon intensity. This highlights that even growth-stimulating fiscal policies contribute to emissions lock-in unless accompanied by structural reductions in carbon intensity, and raises the need to question growth targets themselves.

Usually, the goals to limit global warming are set in terms of total cumulative $CO_2$ emissions $\mathcal{E}$~\cite{seneviratne_allowable_2016, friedlingstein_persistent_2014, peters_challenge_2013, le_quere_trends_2009, friedlingstein_update_2010}, where a given scenario corresponds to a maximal amount of $CO_2$ emitted over a time domain. We can compare the data on cumulative $CO_2$ emissions with the model's expected emissions and obtain a satisfactory match, as shown in Fig.~\ref{fig:fig1}B. We can then use the model to predict emissions for the following years, as shown in Figs. \ref{fig:fig3}A and B, depending on the leverage (used to maintain GDP growth) or growth rate $W$. We observe that cumulative emissions increase with both leverage and growth rate. Using the temperature anomaly as proportional to $CO_2$ emissions~\cite{seneviratne_allowable_2016, friedlingstein_persistent_2014}, we can deduce that the temperature would rise accordingly. However, our results depend on the specific economies of the countries under study and do not capture cumulative emissions on a global scale, nor do they account for global changes in consumption habits.

\subsection*{Study of the solvency probability}

Uncertainties in economy cannot be ignored as they can lead to bankruptcy and a lack of solvency. We can relate the probability of bankruptcy to the probability that a biased random walk is smaller than a threshold at payback time which is fixed $t_p$~\cite{hod_survival_2020, hod_marginally_2020, majumdar_universal_2010}. We emphasize that bankruptcy does not correspond to the existence of a time where the debt is larger than the gross capital (or first passage time), but to the fact that at the time of payback $t_p$, the debt is larger than the gross capital. The solvency probability is a related quantity, which represents the likelihood that debt can be paid back before capital depletion. 

\paragraph*{Strategy A}

We find that the probability of solvency depends mainly on the difference between the long-term growth rate and the logarithm of the interest rate. Interestingly, three regimes are possible depending on the sign of this difference, leading to different optimal strategies. If the long-term growth rate is strictly larger than the logarithm of the interest rate, the probability of solvency goes to $1$ as payback time increases, while the opposite trend is observed (solvency probability going to $0$) if the long-term growth is strictly lower than the logarithm of the interest rate. Thus, there is no pressure to payback in the first case, while the payback should occur as early as possible in the second case to avoid bankruptcy. In addition, there is a limiting case, where the solvency probability approaches $1/2$ as the payback time tends to infinity.

\paragraph*{Strategy B}

In the case of strategy B, unlike the case of strategy A, the solvency probability cannot be written in a closed form. We derive instead a recursive equation for the net capital from which we deduce that this solvency probability is a decreasing function of leverage provided that the intrinsic growth rate is lower than the interest rate and leverage is sufficiently close to $1$.  We observe this trend in Fig.~\ref{fig:fig4} and provide details of the calculations in Methods section and in the Supplementary Information.

Typically, solvency is lower under strategy B because a higher risk must be taken to sustain a higher growth rate. For $L_{\tau} \sim 1$, this solvency probability (strategy B) is smaller than for strategy A, since the investor relies more on leverage in strategy B than in A. 

%\subsubsection*{Effect of leverage on solvency (strategy B)}

We then examine the impact of leverage and payback time on the probability of solvency. As expected, we observe that solvency probability generally decreases with both leverage and payback time, as illustrated in Fig.~\ref{fig:fig4}. In other words, debt leads to short-term GDP gains, even if it results in an increased long-term risk and enhanced cumulative carbon emissions, especially when the debt grows faster than GDP.
In SI 2.4 and 2.5, we further explore solvency in both models.
Notably, we find that high leverage strategies are optimal when the objective is the maximization of the growth rate for a given level of risk as measured by the solvency probability. However, these  strategies are only optimal because environmental costs are ignored. In practice, high leverage strategies typically lead to large carbon emissions, particularly for carbon intensive economies, as discussed above.

\begin{figure*}
    \centering
    \includegraphics[width=\linewidth]{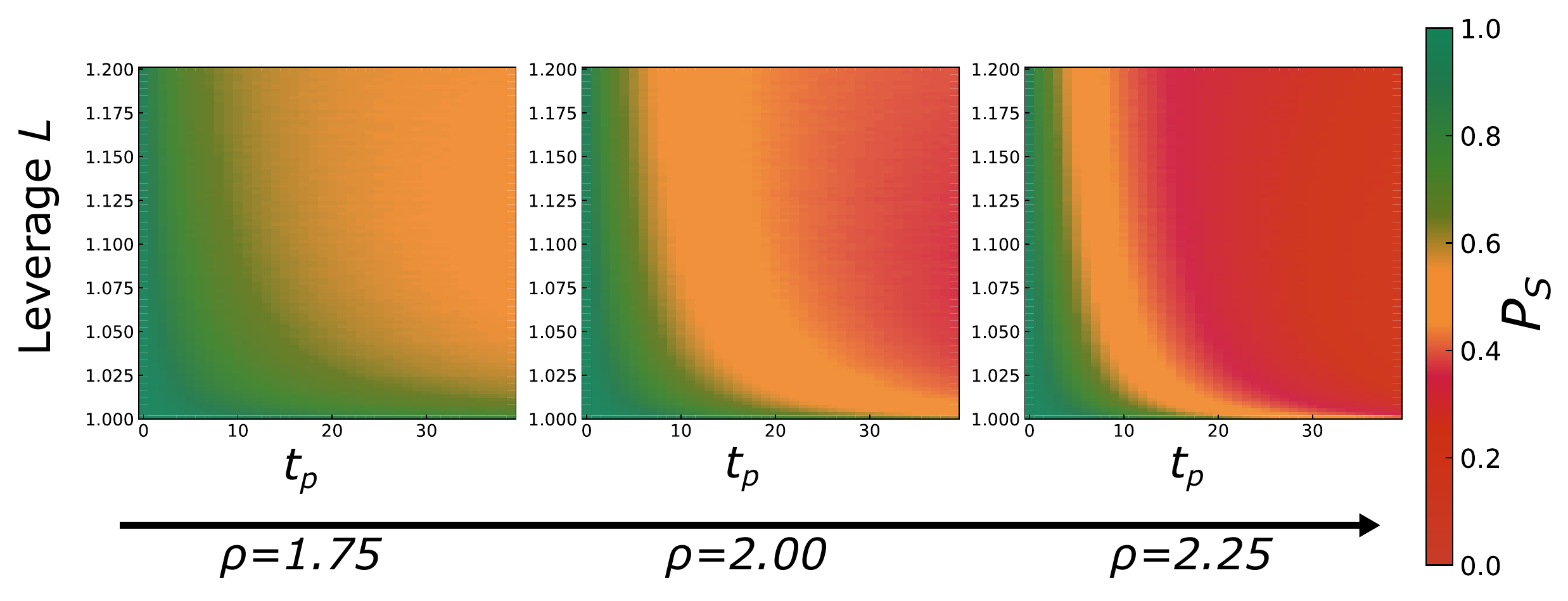}
    \caption{Illustration of trade-off between solvency and payback time. Solvency probability $P_S$ is shown for model B with the color scale as function of payback time $t_p$ and interest rate $\rho-1$. Solvency becomes less likely with time in this model whenever $\avg{\gamma}<\rho$ for all leverages $L$, {\it i.e.} when debt increases faster than the capital. This means that this strategy is not viable on long timescales, and is just a way to increase short-term GDP (resulting in increased carbon emissions).}
    \label{fig:fig4}
\end{figure*}

\section*{Discussion}

Our analysis reveals that debt-financed growth structurally amplifies emissions. Specifically, the leverage $L$ boosts GDP $C_\tau$, while the slow decline in carbon intensity $I(\tau)$ leads to cumulative emissions that remain nearly proportional to cumulative GDP. This confirms previous studies~\cite{tucker_carbon_1995},~\cite{hickel_is_2020},~\cite{lamb_transitions_2014}. 
By separating the growth due to deficits from the intrinsic growth, we have confirmed that the economic growth of several countries is closely tied to deficits. 

A second important observation is that although the intensity of carbon $I(\tau)$ generally declines due to technological improvements and efforts to develop clean energy, increasing the leverage $L$ always amplifies GDP $(C_\tau)$, compensating technological gains in clean energy production. 
As a result, we predict that total emissions $\epsilon(t)$ will continue to increase in the coming decades.

We calibrated the model over multi-decade windows using observed macroeconomic data, extracting each country’s effective leverage $L$ (the average ratio of deficit-financed spending to GDP), the intrinsic growth rate $\gamma$ (the underlying growth rate absent new borrowing), and an interest rate $\rho-1$.
This data-driven approach is not a free fit, but rather implements recorded deficits, public debt, and GDP to set model parameters. Therefore, the model's predictions can be directly compared to the actual outcomes. 

To our knowledge, this is the first model to explicitly link fiscal leverage with cumulative CO$_2$ emissions and to calibrate it using multi-decade country-level data. Indeed, economic growth is typically treated as exogenous or as the result of a production function in Integrated Assessment Models (IAMs), such as DICE, REMIND, or MESSAGEix, without explicitly modelling credit dynamics~\cite{huppmann_messageix_2019, nordhaus_integrated_2015}.  Similarly, the majority of climate-finance studies focus on transition risks or stranded assets~\cite{battiston_accounting_2021}, rather than quantifying the structural amplification of GDP growth and emissions by leverage.  Our method is similar to macro-energy analyses that connect debt and energy consumption~\cite{hagens_economics_2020, king_integrated_2019}. However, it goes further by calibrating leverage $L$, intrinsic growth $\gamma$, and interest rate $\rho-1$ based on historical data. This allows for the direct prediction and validation of cumulative emissions trajectories.
For all four considered countries, the model accurately tracks the cumulative CO$_2$ emissions per capita alongside economic growth (Fig.~\ref{fig:fig2}B and SI), confirming that cumulative emissions are closely coupled with cumulative GDP, thus resulting in a persistent carbon lock-in effect. We can observe that, despite differences in decarbonization efforts, each additional unit of GDP has historically carried a largely invariant carbon cost over the long run (within the calibration period), carrying near-proportional relationships between total economic output and total emissions. 
For example, using a time-declining carbon intensity $I(t)$ for the US, the integration of our model from 1990 to 2020 yields cumulative per capita CO$_2$ emissions of the order $7\times10^5$ kg, in good agreement with the observed value of $\sim6\times10^5$ kg.
A similar close agreement is found for China, France, and Denmark (Fig.~\ref{fig:fig2}B), as evidenced by the overlapping model trajectories and historical data points. This level of precision indicates that our leveraged growth model captures a first-order causal driver of emissions: debt-fuelled economic expansion. Hence, it can provide a partial explanation for why emissions have continued to increase in line with commodity production, despite gains in energy efficiency in many sectors.
The differences between countries in economic structure and debt dynamics are reflected in the calibrated parameters and outcomes.
We find that the United States is heavily dependent on deficit spending to sustain its growth. Indeed, since 1980, its intrinsic growth rate has averaged to only one percent per year, with substantial variability. In other words, most of the US GDP growth since 1980 can be attributed to credit expansion (public deficit and debt accrual) rather than to productivity gains~\cite{magazzino_economic_2012}. Our calibration leads to an average leverage slightly above 1 for the US, rising to $L\approx1.06$ in 2023 amid large federal spending. This credit-driven growth is reflected in the model’s ability in reproducing the persistent rise in cumulative US CO$_2$ emissions at the end of the 90s until just before the global financial crisis of 2008 (GFC)~\cite{feng_drivers_2015}. 

Quantitatively, the correspondence between modeled and observed cumulative CO$_2$ is such that a simple linear regression of one against the other would yield a slope near 1, {\it i.e.} cumulative emissions are almost a linear function of cumulative GDP (see Fig.~\ref{fig:fig2}C). The scatter of points around that line is small, thus reflecting primarily year-to-year economic variability rather than structural error in the model. 
This accordance is relevant, given that we did not calibrate the model by fitting it to emissions data, but rather derived emissions from independently calibrated economic parameters and an exogenous decline in carbon intensity. 
The observed results reveal that the coupling between economic output and emissions (illustrated by the Kaya identity,~\cite{tavakoli_how_2017}) is grounded in the macro-financial structural dynamics of growth and investment. 

An additional insight from our framework is that debt-driven growth imposes financial risks and constraints. 
Our stochastic analysis of the solvency probability (Fig.~\ref{fig:fig4} and SI) shows that when the intrinsic growth rate $\langle \gamma \rangle$ falls below the interest rate $\rho-1$, the probability of long-term solvency decays towards zero, regardless of leverage $L$. 
In other words, maintaining growth through excessive borrowing ultimately becomes unsustainable and carries an endogenous risk of default. 
This result imposes a natural constraint on fiscal stimulus strategies. Even before critical climate transitions are reached, macro-financial dynamics may trigger deleveraging and systemic destabilizations.  
From a control theory perspective, this identifies an optimal leverage level on the Pareto front, beyond which additional credit raises default risk more than it supports sustainable growth~\cite{forouli_multiple-uncertainty_2020,everall_pareto_2025,dinis_phase_2020}. 

In this sense, the model provides a quantitative link between carbon debt and financial debt: high-leverage, high-emission growth paths not only increase cumulative CO$_2$ emissions but also increase the risk of abrupt macroeconomic adjustments if interest rates rise or economic growth slows. 
Policies that are based on debt-financed growth should be assessed for their climate implications and the risk they pose of triggering financial instability and default. 
 %The predictive capacity of IAMs could be enhanced by incorporating this feedback, which would enable assessments that account for both credit cycle risks and biophysical constraints.

 One key limitation of the model is its tight coupling of public debt and GDP. We assume that each country’s government borrows a fixed fraction of GDP each year (the leverage $L$ above 1 corresponds to a primary deficit ratio) and that this borrowing directly and immediately adds to that year’s GDP. In reality, the relationship between debt and growth is more complex: private credit creation (by banks and firms) can also drive variability, and not all deficit spending translates into GDP in the short run (in fact, the fiscal multiplier can vary). Our model effectively treats deficit-financed spending as equally productive as other spending, whereas inefficient investments or financial sector leakages could reduce the growth payoff of debt. In addition, identifying the primary deficit with new capital investment is an oversimplification. In fact, governments may borrow for many reasons, and not all borrowing fuels productive capacity.
%China is an example where much debt is funnelled through state-owned enterprises rather than the central budget, meaning our use of fiscal deficit data might underestimate the true leverage acting on the economy.
%Conversely, Denmark’s small deficits (even surpluses) do not mean the private sector had no debt-fuelled activity; they indicate that the public sector did not contribute to it. 
%Ask B. for opinion 
A more advanced model would require integrating both public and private debt dynamics and the ability to distinguish between productive investment and pure demand stimulus. Such a complementary analysis of private debt could be conducted, as done previously with macroeconomic models of financial instability \cite{keen_finance_1995, giraud_macrodynamics_2023}.

%i cut it add if asked
%Furthermore, our framework currently presupposes that countries do not engage in financial interactions; therefore, the model does not account for any cross-border capital flow or trade imbalances. In reality, another country’s surplus can finance one country’s debt, and international credit conditions can influence domestic leverage. Incorporating an open-economy dimension could be a significant extension, as global financial cycles may synchronize or constrain the leverage available across multiple countries simultaneously.

Another important limitation is the model's lack of sectoral differentiation in its treatment of emissions. We model each country as a single homogeneous production unit, with a single aggregate carbon intensity of output, in a manner similar to the Nordhaus model~\cite{nordhaus_optimal_1992, nordhaus_dicemodel_1992, nordhaus_climate_2017, nordhaus_revisiting_2017}. 
This implies that we do not account for changes in the composition of GDP from high-carbon sectors (such as heavy industry) to low-carbon sectors (such as services). 
For example, within a country, some industries can rapidly decarbonize (such as renewables in power generation), while others lag behind (such as air travel or defense)~\cite{ben_youssef_relationships_2023, dong_chinas_2021}. Our one-sector approach averages these dynamics out. Additionally, we assume that the same carbon intensity function $I(t)$ applies regardless of the amount produced, thereby ignoring any nonlinear feedback (such as resource constraints, climate risks, or price effects) that may occur. In reality, a large increase in credit use (as in China’s case) can temporarily increase carbon intensity by increasing investment in construction and infrastructure. Our model partially reflects this by calibrating $I(t)$ piecewise (using historical data to set $I_0$ and a decline rate $\eta$).
%however, as for now, it cannot predict endogenous slower decarbonization during credit booms, faster decarbonization during recessions, or policy-driven transitions. 
China's financial system structure may partially explain the apparent decoupling between GDP growth and territorial CO$_2$ emissions. Unlike market-based systems, in which credit allocation is determined by profit expectations, China's state-dominated banking sector sets lending priorities based on policy objectives. A large-scale financing of renewables, electrified transport, and grid infrastructure is facilitated by the government's determination of which sectors are eligible for credit. According to the model, this centralised allocation temporarily reduces the leverage–emissions gap by increasing the effective decarbonization rate $\beta$ in comparison to the growth rate $\gamma$. 
%IMPORTANT HYSTERESIS - SHORT TERM MEMORY OF PAST GROWTH
%Nevertheless, cumulative intensity $I_\tau$ remains elevated from earlier carbon-intensive growth, resulting in a partial and reversible decoupling (this scale is in the context of long-term carbon lock-in).
%%if the codes are correct, add this
%In our model, China's apparent decoupling can be interpreted as a high-growth, short-memory regime: the rapid, credit-driven expansion (large $W$) exponentially weights recent, cleaner years, causing cumulative and instantaneous intensities to converge, resulting in the appearance of decoupling.  While high, steady growth can make cumulative and instantaneous intensities converge, this regime is intrinsically fragile. The convergence arises from exponential weighting rather than structural efficiency gains. If growth slows, for instance, due to a recession, the cumulative memory of past emissions would re-emerge, revealing a large residual “carbon debt.” 

Denmark presents another informative and distinctive boundary case for our framework, owing to its limited public borrowing and persistent fiscal surpluses.  As illustrated in the Supplementary Information, Denmark's territorial emissions have decreased, while its consumption-based emissions have remained relatively constant, indicating that no actual decoupling has occurred.  This discrepancy suggests that fiscal restraint alone is insufficient to guarantee a reduced carbon footprint when emissions are outsourced through trade.
In the Supplementary Information, we compare territorial and consumption-based estimates (Fig.~S2) and find that offshoring of emissions can significantly alter apparent decoupling trends (a phenomenon also referred to as carbon leakage~\cite{jakob_why_2021}). Despite their high consumption-based footprint, countries that relocate carbon-intensive production abroad appear to decrease their emissions-to-GDP ratio~\cite{matthews_proportionality_2009}. To address this issue, network-based allocation methods developed by Van den Ende et al. (2025) offer a rigorous approach to quantify responsibility for emissions along global value chains, which complements territorial accounting~\cite{van_den_ende_network-based_2025}. 

%A more comprehensive representation of trade-embedded emissions and globalised carbon fluxes would be possible through the integration of multi-regional input–output (MRIO) data into the model. 
 
 It's worth noticing that our analysis does not currently include private credit and investment flows, which can be significant drivers of emissions and capital formation~\cite{popescu_investment_2024}.  The current representation of the leverage term $L_t$ is based on public fiscal deficits, which account only for the government's contribution to output growth through borrowing. Nevertheless, alternative data sources, such as total credit to the non-financial sector or capital market issuance, could be used to illustrate the intensification of growth through private investment flows.
 This expansion would enable the model to quantitatively describe mechanisms, such as those specifically identified in recent analyses on institutional investors and insurers, that persist in financing the expansion of fossil fuels despite climate commitments~\cite{reclaim_finance_insurance_2024},~\cite{schreiber_climate_2020}.

%Such flows effectively act as a shadow leverage channel: they expand economic activity and lock in carbon-intensive capital even without public borrowing. 
%https://inis.iaea.org/records/jd2pc-7dw16
%https://reclaimfinance.org/site/en/2024/12/10/insurance-scorecard-2024-insurers-fuelling-an-uninsurable-world/

%Our model enables a quantitative evaluation of the impact of alternative credit allocation strategies on mitigation and adaptation outcomes by connecting credit creation, growth, and cumulative emissions. 
%A future research perspective could be to develop a policy-relevant tool that identifies fiscal and financial policies aligned with net-zero pathways. In a similar way to~\cite{andreoni_policy_2025, battiston_accounting_2021}, combining detailed policy modeling and emissions accounting, our method could be used to systematically assess whether deficit spending and credit allocation are consistent with the remaining carbon budget, providing a connection between macroeconomic policy and climate targets suitable for integration into future IPCC scenario assessments. 
%tbc 
%missing part on China/bankruptcy/risk/optimality 

Financial debt and carbon debt are two sides of the same anticipatory system: the former brings future wealth forward, while the latter consumes the energy of the future. Our stochastic credit-risk model, when scaled up to the macroeconomic level, shows that there are both physical and financial limits to debt-driven growth. 
It is an important message because the idea that there are limits to growth, even when allowing for innovation, is not widely accepted among economists \cite{susskind_growth_2024}. Here, we show that as long as credit rests on the promise of material expansion, each euro borrowed creates a double debt — a financial and a climatic one. The great challenge of the twenty-first century will be to decouple prosperity from the credit lever, so that the growth of well-being is no longer tied to the growth of carbon.

\section*{Methods}

\paragraph{Definition of the stochastic credit-risk model }
%All variables are expressed in consistent units: $Y$ in USD (constant 2015), $C$ in tCO$_2$, and $\epsilon$ in tCO$_2$/USD.
Our stochastic credit-risk model is defined by the following dynamical equations for the capital $C_\tau$ and the debt $D_\tau$ at time $\tau$: 
\begin{align}
    C_0 &= B_0 + C_{\text{ini}} =LC_{\text{ini}},
    \\
    C_{\tau + 1} &= \gamma_{\tau}C_{\tau} + B_{\tau+1} - \left(\sum_{\tau'=0}^{\tau}\delta_{\tau,t_p(\tau')}\right)D_{\tau+1},
    \\ D_0 &= B_0 = (L-1)C_{\text{ini}},
    \\D_{\tau} &= \sum_{\tau'=0}^{\tau-1} \Theta(t_p(\tau') - \tau)\rho_{\tau'}^{\tau - \tau'}B_{\tau'},
\end{align}
where in the first equation, $C_{\text{ini}}$ is the amount invested initially by the investor, and $B_0$ is the initial amount borrowed. The leverage is $L=(B_0+C_{\text{ini}})/C_{\text{ini}}$. 
The second equation describes the growth of the capital from time $\tau$ to time $\tau+1$. 
The model uses discrete time, as in Kelly's original model. However, extending it to continuous time is straightforward. At each time step, the capital is multiplied by $\gamma_{\tau}$, which accounts for inflation \cite{azoulay_chapitre_2023}. This parameter represents the stochastic instantaneous growth rate at time $\tau$. The evolution of the debt is described by the last two equations.  The interest rate $\rho_{\tau}-1$ is the factor by which the amount borrowed is multiplied at each time step, in other words, the debt due to the amount borrowed at time $\tau'$, is multiplied by $\rho_{\tau'}^{\tau-\tau'}$ if it has not been paid back at time $\tau$. In general, the interest rate is larger than $0$ ($\rho > 1$), so that the debt keeps increasing, and the borrower ends up paying more than the amount borrowed. 
%Then the gross capital $C_{\tau+1}$, and debt $D_{\tau + 1}$ after one step are (until bankruptcy occurs if $C_{\tau+1}\leq 0$):
%(in a classical Kelly's model for horse races this factor would equal the product of the bets and odds on the winning horse~\cite{dinis_phase_2020, cavallero_trade-off_2025}). 
%We define the average and standard deviation of $\gamma_{\tau}$, respectively $W$ and $\sigma_{\ln(\gamma_{\tau})}$. 
%The first term in the first equation represents the contribution of stochastic growth at time $\tau$: 
If $\gamma_{\tau}>1$, the gross capital increases provided that there is no borrowing or payback at this time. The second term represents the amount borrowed at time $\tau$; it can be $0$ if the borrower does not borrow at this step, or a positive value. The third term represents the possible payback of the debt (debt clearing) at time $\tau+1$, and will either be equal to $0$ (no payback), or to $D_{\tau+1}$. 

In the second and fourth equation, $t_p(\tau')$ indicates the time when the amount borrowed at $\tau'$ is paid back. In the second equation, 
if a payback happens at time $\tau$, $\exists \tau', \tau = t_p(\tau')$, and thus $\left(\sum_{\tau'}\delta_{\tau,t_p(\tau')}\right)=1$, then the capital is reduced by the service of the debt, otherwise $\left(\sum_{\tau'}\delta_{\tau,t_p(\tau')}\right)=0$ and there is no such reduction.
In the last equation, the Heaviside function $\Theta$ function allows us to know if the payback has occurred at time $\tau$ (indeed $\Theta(t_p(\tau') - \tau)=0$ if $\tau>t_p$, meaning that the debt is erased after it has been payed back). 

\paragraph{Strategy A}

We assume that the investor initially borrows money ($B_0>0$) but never borrows afterwards, $\forall \tau \geq 1, B_{\tau} = 0$. We refer to this strategy as strategy A, which is akin to an individual taking out a loan to buy a house. In practice, the investor can borrow $B_0$ and invest $C_{\text{ini}}$ from its personal capital, and the leverage is then $L=(B_0+C_{\text{ini}})/C_{\text{ini}}$. In this case, for $\tau < t_p$, the gross capital evolves as in the standard Kelly's model with an initial investment of $B_0+C_{\text{ini}}$, and the debt evolves with an interest rate $\rho-1$. At the time of payback, the capital can become negative, which could not happen within the purely multiplicative dynamics of Kelly's model. Bankruptcy occurs if $C_{t_p}$ is negative, or equivalently if 
\begin{equation}
\prod_{\tau'\leq t_p} \gamma_{\tau'} < 1 -\frac{1}{L},
\end{equation}
meaning that the larger $L$ is, the more likely the bankruptcy. Under the assumption of no bankruptcy, we can also define the long-term growth rate $W = \avgE{\ln\left(\gamma\right)}$.  Note that due to the concavity of the logarithm, we have generally for any probability distribution of $\gamma$, 
$\ln(E) \geq W$,
with $E=\avgE{\gamma}$. However, there is a non-zero probability of hitting bankruptcy when the debt is paid back (at $t_p$). 

\paragraph{Strategy B }

Within strategy B, the investor borrows to an external actor at each time step with a time dependent leverage $L_{\tau}$. The dynamics is then (except for the time of payback):
\begin{equation}
\begin{split}
    C_{\tau+1} &= \left(\gamma_{\tau} + (L_{\tau}-1)\right)C_{\tau},
    \\ D_{\tau+1} &= \rho D_{\tau} + (L_{\tau}-1)C_{\tau}.
\end{split}
\end{equation}

Now, in addition to the long-term growth rate $W$ defined as before, one can also consider the short-term growth rate $\omega(t_p)$ (defined in SI section 1.1). We explicitly separate the value linked to production from the value added through borrowing, in the stochastic \textit{intrinsic growth rate}:
\begin{equation}
    r(\tau) = \frac{C_{\tau + 1}-C_{\tau}}{C_{\tau}} - (L_{\tau}-1) = \gamma_{\tau} - 1.
\end{equation}

\paragraph{Solvency in Strategy A}

The solvency probability $P_S(t_p)$ for a fixed payback time $t_p$ can be computed explicitly in this case. To do so, we introduce the stochastic variable $X_{\tau'} = \ln(\gamma_{\tau'} / \rho)$, which represents increments of the long-term growth rate. This variable has mean $m=W-\ln(\rho) $ and variance $\sigma_X^2$.
Then, we show in Supplementary Information that 
\begin{equation}
    P_S(t_p)=\frac{1}{2}\left(1+\text{erf}\left(\frac{t_p m -\ln\left(1-\frac{1}{L}\right)}{\sqrt{2t_p}\sigma_X}\right)\right).
    \label{eq:surv_gauss}
\end{equation}
From this equation, we find that three regimes are possible depending on the sign of $m$. If $m>0$, $P_S(t_p\to\infty) \to 1$, solvency is guaranteed, and there is thus little pressure to pay back the debt. If $m<0$ instead, $P_S(t_p\to\infty) \to 0$ and the convergence is faster as $\sigma_X$ decreases, but in this case, the best course of action is to pay back the debt as soon as possible to avoid bankruptcy.
The case $m=0$ is a limiting case where $P_S(t_p\to\infty) = \frac{1}{2}$.

\paragraph{Solvency in Strategy B}

The expression for the solvency probability in this case is
\begin{equation}
\label{eq:solvency_proba}
\begin{split}
    P_S(t_p) = \mathbb{P}\left(X_{t_p}\geq 1-\frac{1}{L}\right),
\end{split}
\end{equation}
with $X_{t_p}$ now defined by $X_{t_p} = \Pi_{t_p} -(L-1)\sum_{t\leq t_p - 1} \Pi_t/\rho $ with $\Pi_t = \prod_{1\leq\tau'\leq t} \left(\gamma_{\tau'} + (L-1)\right)/\rho$. We identify two regimes in this case, which are controlled  by the sign of $\avg{\gamma}  -\rho$. Under the assumption that fluctuations in the economy are not too large, we expect the solvency probability to be a decreasing function of time when $\avg{\gamma} < \rho$, whereas the opposite is expected for $\avg{\gamma} > \rho$. Thus, we recover a well-known condition that sets the stability of public debt with respect to GDP~\cite{azoulay_chapitre_2023}. From this, we conclude that debt tends to rise faster on average than capital as long as the \textit{intrinsic growth rate} is smaller than the interest rate $\rho-1$. 

\paragraph{Data analysis}

We have extracted from economic data of four countries (China, US, Denmark and France) in the time frame $1998-2022$, parameters of the model
as shown in Table \ref{tab:measur_parameters}. 

\begin{table}[ht]
    \centering
    \begin{tabular}{|c|c|c|c|c|}
         \hline & China & US & Denmark & France 
         \\ \hline Average leverage $\avg{L_{\tau}}$ \hspace{-0.23cm} & \hspace{-0.23cm}$1.025 \pm 0.023$ \hspace{-0.23cm}& \hspace{-0.23cm}$1.051 \pm 0.040$ \hspace{-0.23cm}&\hspace{-0.23cm} $0.997 \pm 0.029$ \hspace{-0.23cm}&\hspace{-0.23cm} $1.040 \pm 0.018$\hspace{-0.23cm}
         
         \\ \hline Average interest rate $\avg{\rho_{\tau}}-1$ \hspace{-0.23cm} &\hspace{-0.23cm} $0.131 \pm 0.101$ \hspace{-0.23cm}&\hspace{-0.23cm} $0.013 \pm 0.051$ \hspace{-0.23cm}&\hspace{-0.23cm} $0.018 \pm 0.237$ \hspace{-0.23cm}&\hspace{-0.23cm} $0.008 \pm 0.088$ \hspace{-0.23cm}
         
         \\ \hline Average long-term growth rate $W$ \hspace{-0.23cm} &\hspace{-0.23cm} $0.088 \pm 0.086$ \hspace{-0.23cm}&\hspace{-0.23cm} $-0.018 \pm 0.059$ \hspace{-0.23cm}&\hspace{-0.23cm} $0.032 \pm 0.089$ \hspace{-0.23cm}&\hspace{-0.23cm} $-0.020 \pm 0.090$ \hspace{-0.23cm}
         \\ \hline Average intrinsic growth rate $\avg{\gamma_{\tau}} - 1$ \hspace{-0.23cm} &\hspace{-0.23cm} $0.096 \pm 0.095$ \hspace{-0.23cm}&\hspace{-0.23cm} $-.017 \pm 0.056$ \hspace{-0.23cm}&\hspace{-0.23cm} $0.036 \pm 0.091$ \hspace{-0.23cm}&\hspace{-0.23cm} $0.017 \pm 0.088$ \hspace{-0.23cm}
         \\ \hline
    \end{tabular}
    \caption{Estimations of key parameters of the model from economic data of GDP and deficits of four different economies.}
    \label{tab:measur_parameters}
\end{table}

\section*{Acknowledgments}

We are grateful to Eric Galbraith, Julia Steinberger, Jeff Althouse, and André de Palma for their insightful feedback and discussions, which greatly strengthened this work. S.M. acknowledges support from the 2025 FIR–PRI (Best Research Grant) for her PhD project.

\appendix

\section{A stochastic macroeconomic credit-risk model}

In this section, we provide more details on the stochastic credit-risk model studied in the section on Methods in the main text. There are two main variants of the model depending whether there is a single load (strategy A) or periodic borrowing (model B).
Then, we also explain how this model relates to known macroeconomic models.
In the next section, we provide more details on the data analysis and on the evaluation of carbon emissions.

\subsection{Single loan model (strategy A)}
For strategy A, we assume that the gambler initially borrows money ($B_0>0$) but never borrows afterwards, $\forall \tau \geq 1, B_{\tau} = 0$. This strategy is more like that of an individual taking out a loan to buy a house. In practice, the gambler can borrow $B_0$ and invest $C_{\text{ini}}$ from its personal capital, and a key quantity will be the leverage $L=(B_0+C_{\text{ini}})/C_{\text{ini}}$. In this case, for $\tau < t_p$, the gross capital evolves as in the standard Kelly's model with an initial investment of $B_0+C_{\text{ini}}$, and the debt evolves with an interest rate $\rho-1$

\begin{equation}
\begin{split}
    C_{\tau} &= (B_0+C_{\text{ini}}) \prod_{1\leq\tau'\leq\tau} \gamma_{\tau'} = C_{\text{ini}} L \prod_{1\leq\tau'\leq\tau} \gamma_{\tau'}.
    \\ D_{\tau} &= B_0 \rho^{\tau} = C_{\text{ini}}(L-1) \rho^{\tau}
\end{split}
\end{equation}

 Then at the time of payback, for $\tau = t_p$, we find:
\begin{equation}
\begin{split}
    C_{t_p} &= B_0 \left(\prod_{1\leq\tau'\leq t_p} \gamma_{\tau'}- \rho^{t_p}\right) + C_{\text{ini}} \prod_{\tau'\leq t_p} \gamma_{\tau'}
    \\ &=  C_{\text{ini}}\left[L \left(\prod_{1\leq\tau'\leq t_p} \gamma_{\tau'}-\rho^{t_p}\right) +  \rho^{t_p}\right]
\end{split}    
\end{equation}

Bankruptcy occurs if $C_{t_p}$ is negative or equivalently
\begin{equation}
    \prod_{\tau'\leq t_p} \frac{\gamma_{\tau'}}{\rho} < 1 -\frac{1}{L},
\end{equation}

 meaning that the larger $L$ is, the more bankruptcy is likely to occur. In average over the joint probability $p(x_1, ..., x_{t_p})$, using the fact that the races are independent, we find:
\begin{equation}
\begin{split}
    \avg{C_{t_p}}&=C_{\text{ini}}\left[L\left(\sum_{x_1, ..., x_{t_p}}\mathbb{P}(x_1, ...x_{t_p})\prod_{i=1}^{t_p}\gamma_{x_i}-\rho^{t_p}\right)+\rho^{t_p}\right]
    \\&=C_{\text{ini}}\left[L\left(\sum_{x_1, ..., x_{t_p}}\prod_{i=1}^{t_p}\mathbb{P}(x_i)\gamma_{x_i}-\rho^{t_p}\right)+\rho^{t_p}\right]
    \\&=C_{\text{ini}}\left[L\left(\prod_{i=1}^{t_p}\sum_{x_i}\mathbb{P}(x_i)\gamma_{x_i}-\rho^{t_p}\right)+\rho^{t_p}\right]
    \\&=C_{\text{ini}}\left[L\left(E^{t_p}-\rho^{t_p}\right)+\rho^{t_p}\right],
\end{split}
\end{equation}   
 where $E=\mathbb{E}(\gamma_x)$ is the average earning after each race (in the following we will consider the case $E>1$). We can also define the average short-term growth rate
\begin{equation}
\label{eq:w_t_p}
    \omega(t_p) = \frac{1}{t_p}\ln\left(\frac{\avg{C_{t_p}}}{LC_{\text{ini}}}\right) = \frac{1}{t_p}\ln\left(E^{t_p}-(1-\frac{1}{L})\rho^{t_p}\right),
\end{equation}

which is the average growth rate of capital relative to the initial capital $LC_{\text{ini}}$ just after the debt is repaid, assuming no bankruptcy. This way we can write that $C_{t_p} = LC_{\text{ini}} \exp(\omega(t_p) t_p)$. The larger this quantity is, the faster the capital will grow after payback. If the gambler is thinking in short-terms, this is a quantity to maximize. Interestingly we can bound this quantity (because $(1-1/L)\rho^{t_p} > 0$) : 
\begin{equation}
    \forall t_p, \ln(E) \geq \omega(t_p)
\end{equation}

If bankruptcy does occur, the game stops. Otherwise, the gambler can keep playing, and for $\tau>t_p$:
\begin{equation}
    C_{\tau} = C_{\text{ini}}\left[  L\left(\prod_{\tau'\leq t_p} \gamma_{\tau'} - \rho^{t_p}\right) + \rho^{t_p}\right]\prod_{t_p<\tau'\leq \tau} \gamma_{\tau'}.
\end{equation}

 From this we see that the gross capital is well defined for $(1/\rho^{t_p})\prod_{\tau'\leq t_p}\gamma_{\tau'} > 1 - 1/L$. Under this assumption (no bankruptcy), the growth rate conditioned on solvency is defined as:
\begin{equation}
\begin{split}
    W &= \lim_{\tau\to\infty}\frac{\ln(C_{\tau})}{\tau} = \lim_{\tau\to\infty} \frac{1}{\tau}\left(\ln\left(\prod_{t_p<\tau'\leq \tau} \gamma_{\tau'}\right) + \ln\left(C_{\text{ini}}\left[  L\left(\prod_{\tau'\leq t_p} \gamma_{\tau'} - \rho^{t_p}\right) + \rho^{t_p}\right]\right)\right)
    \\ &= \lim_{\tau\to\infty} \frac{1}{\tau}\left(\ln\left(\prod_{t_p<\tau'\leq \tau} \gamma_{\tau'}\right)\right)
    \\& = \avgE{\ln(\gamma_x)},
\end{split}
\end{equation}

 where the equivalence results from the weak law of large numbers, and the fact that the second term is a constant with respect to the time $\tau$ (thus disappears when multiplied by $1/\tau$ for large times). Note that due to the concavity of $\ln$, we always have (independent of the probability distribution $p_x$):
\begin{equation}
    \ln(E) \geq W
\end{equation}

However, there is a non-zero probability of hitting bankruptcy when the debt is paid back (at $t_p$). The probability of hitting bankruptcy is:
\begin{equation}
    P_B(t_p) = \mathbb{P}\left(\prod_{\tau'\leq t_p}\frac{\gamma_{\tau'}}{\rho}\leq 1-\frac{1}{L}\right) = \mathbb{P}\left(\sum_{\tau'\leq t_p}\ln(\frac{\gamma_{\tau'}}{\rho}) \leq \ln\left(1-\frac{1}{L}\right)\right).
\end{equation}

 Within Kelly's model, and with a strategy corresponding to Kelly's optimum $b_x = p_x$, we obtain stochastic evolutions of the gross capital on Fig.1 from the main text. We observe that some trajectories hit bankruptcy at the time of payback, while others survive and keep increasing, depending on the history of results for $\tau \leq t_p$.

If we fix the probability distribution $p_x=[0.03, 0.20, 0.39, 0.12, 0.26]$ and the inverse odds $r_x = [0.15, 0.39, 0.04, 0.21, 0.21]$ using Kelly's optimal strategy $b_x = p_x$, we plot the simulated solvency probability for different interest rates on Fig.\ref{fig:gaussian_survprob_A} in the limit of large leverage $L=15$. Depending on the value of the interest rate, the solvency probability can either increase (low interest rates) or decrease (large interest rates) with the payback time $t_p$.

\subsection{Periodic borrowing model (strategy B)}

In this section, we assume that the gambler borrows money at each time step, with a leverage $L_{\tau}$ (depending on time a priori), which we refer to as strategy B. In this case, for $\tau < t_p$, the gross capital evolves as in the standard Kelly's model, with a modified stochastic growth rate, and the debt still evolves with an interest rate $\rho-1$. The gambler might want to partially pay back the debt at each step, but we ignore this possibility for now. Before the debt is paid, we have the equations modified in the form

\begin{equation}
\begin{split}
    C_{\tau+1} &= \left(\gamma_{\tau} + (L_{\tau}-1)\right)C_{\tau}
    \\ D_{\tau+1} &= \rho D_{\tau} + (L_{\tau}-1)C_{\tau},
\end{split}
\end{equation}

where the second term in the first equation comes from credit, as well as the second term in the second equation. $G_t$ corresponds to government expenditures, while $T_t$ corresponds to taxation.

We assume that the leverage is constant in time in the following, and justify this assumption on Fig.\ref{fig:primarydef_to_GDP}. Therefore, if we introduce $\tilde{D}_{\tau} = D_{\tau}/\rho^{\tau}$

\begin{equation}
\begin{split}
   C_{\tau} &= C_{\text{ini}} L \prod_{1\leq\tau'\leq\tau} \left(\gamma_{\tau'} + (L-1)\right)
    \\ \tilde{D}_{\tau+1} &= \tilde{D}_{\tau} + \frac{L-1}{\rho^{\tau+1}}C_{\tau},
\end{split}
\end{equation}

and thus, from the second equation

\begin{equation}
    \tilde{D}_{\tau} = \tilde{D}_0 + \sum_{\tau'=0}^{\tau-1}\frac{L-1}{\rho^{\tau+1}}C_{\tau}.
\end{equation}

And finally, we find an expression for the debt

\begin{equation}
    D_{\tau} = \rho^{\tau}\left(D_0 + \sum_{\tau'=0}^{\tau-1}\frac{L-1}{\rho^{\tau+1}}C_{\tau}\right).
\end{equation}

So that we obtain

\begin{equation}
\begin{split}
    C_{\tau} &= C_{\text{ini}} L \prod_{1\leq\tau'\leq\tau} \left(\gamma_{\tau'} + (L-1)\right)
    \\ D_{\tau} & = \rho^{\tau}\left(C_{\text{ini}}(L-1) + \sum_{0\leq t\leq \tau - 1}\frac{L-1}{\rho^{t+1}}C_t\right) = C_{\text{ini}}(L-1)\rho^{\tau}\left(1 + \sum_{0\leq t\leq \tau - 1}\frac{L}{\rho^{t+1}} \prod_{1\leq\tau'\leq t} \left(\gamma_{\tau'} + (L-1)\right)\right).
\end{split}
\end{equation}

 Then at the time of payback, for $\tau = t_p$, we find:
\begin{equation}
\begin{split}
    C_{t_p} &= C_{\text{ini}} L \prod_{1\leq\tau'\leq t_p} \left(\gamma_{\tau'} + (L-1)\right) - C_{\text{ini}}(L-1)\rho^{t_p}\left(1 + \sum_{0\leq t\leq t_p - 1}\frac{L}{\rho^{t+1}} \prod_{1\leq\tau'\leq t} \left(\gamma_{\tau'} + (L-1)\right)\right)
    \\ &= C_{\text{ini}} \rho^{t_p}\left[L\left(\prod_{1\leq\tau'\leq t_p} \frac{1}{\rho}\left(\gamma_{\tau'} + (L-1)\right) - 1-\sum_{0\leq t\leq t_p - 1}\frac{L-1}{\rho} \prod_{1\leq\tau'\leq t} \frac{1}{\rho}\left(\gamma_{\tau'} + (L-1)\right)\right) + 1\right]
\end{split}    
\end{equation}

Therefore, using the same method as previously, we obtain the average

\begin{equation}
    \avg{C_{t_p}}=C_{\text{ini}} \rho^{t_p}\left[L\left(\left(\frac{E+(L-1)}{\rho}\right)^{t_p} - 1-\frac{L-1}{\rho}\frac{1-\left(\frac{E+(L-1)}{\rho}\right)^{t_p}}{1-\frac{E+(L-1)}{\rho}}\right) + 1\right]
\end{equation}

This sets the average short-term growth rate in this case

\begin{equation}
\label{eq:solvency_prob_B}
    \omega(t_p) = \frac{1}{t_p}\ln\left(\rho^{t_p}\left(\left(\frac{E+(L-1)}{\rho}\right)^{t_p} - 1-\frac{L-1}{\rho}\frac{1-\left(\frac{E+(L-1)}{\rho}\right)^{t_p}}{1-\frac{E+(L-1)}{\rho}} + \frac{1}{L}\right)\right)
\end{equation}

This way we can also study the solvency probability for strategy B, which is

\begin{equation}
\begin{split}
    P_B(t_p) &= \mathbb{P}\left(\prod_{1\leq\tau'\leq t_p} \frac{1}{\rho}\left(\gamma_{\tau'} + (L-1)\right) -\sum_{t\leq t_p - 1}\frac{L-1}{\rho} \prod_{0\leq 1\leq\tau'\leq t} \frac{1}{\rho}\left(\gamma_{\tau'} + (L-1)\right) \leq 1-\frac{1}{L}\right).
\end{split}
\end{equation}

\subsection{Relation to public debt models}

In this section, we explicitly relate the model for the public debt of 
\cite{dallery_comptabilite_2023} to the gambling model outlined above. For the public debt, one usually studies debt in terms of the debt-to-GDP ratio. We call $C_{\tau}$ the GDP at time $\tau$, which can be split in different sectors \cite{godley_monetary_2007, dallery_comptabilite_2023}: consumption $U_{\tau}$, investments $I(\tau)$, government spending $G_{\tau}$, exports $X_{\tau}$ and imports $M_{\tau}$ (setting $X_{\tau}-M_{\tau}$ to be the net exports). The primary deficit represents the value added to the public debt (apart from interests), and is split as government spending minus taxation $T_{\tau}$.

\begin{equation}
\begin{split}
    C_{\tau} &= U_{\tau} + I(\tau) + G_{\tau} + (X_{\tau} - M_{\tau})
    \\ D_{\tau+1} &= \rho D_{\tau} + G_{\tau+1} - T_{\tau+1},
\end{split}
\end{equation}

To map this with the previous model requires assuming that the primary deficit $G_{\tau}-T_{\tau}$ is a fraction of the GDP, borrowed to increase the yearly GDP. Even if the mechanism is more complex, and debt usually motivates carbon-intensive activities based on the assumption that taxes will eventually generate surpluses for the government. If we want to identify with our previous model, we can write that

\begin{equation}
    C_{\tau + 1} = \gamma_{\tau} C_{\tau} + G_{\tau+1} - T_{\tau+1}.
\end{equation}

\begin{figure}
    \centering
    \includegraphics[width=\linewidth]{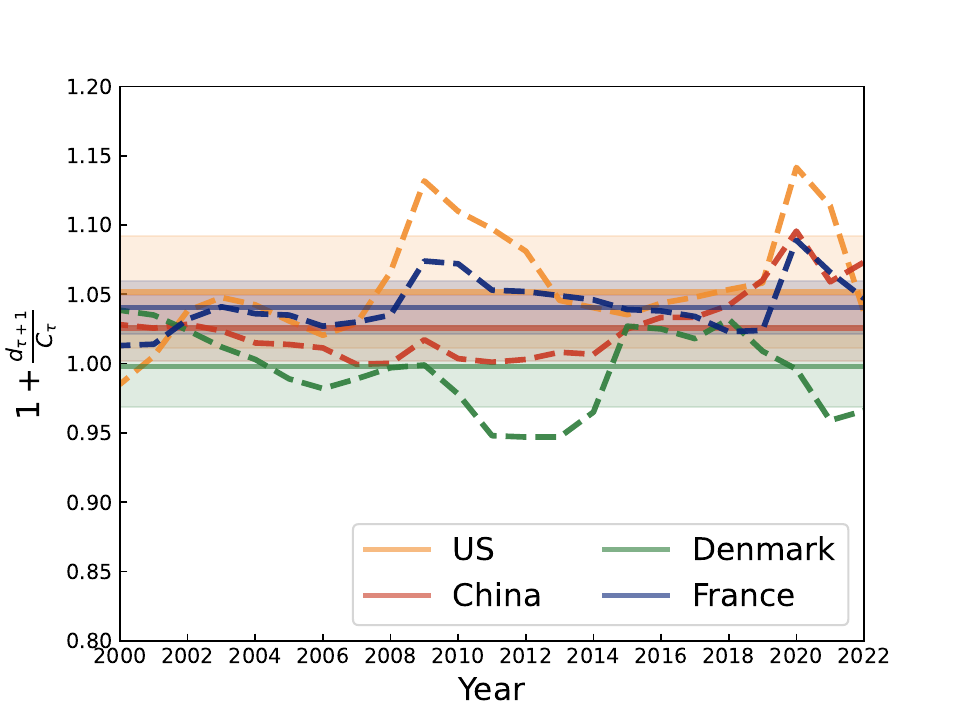}
    \caption{Comparing primary deficit to GDP for different countries. In dotted lines we show the data, in full lines the average values and the colored areas represent standard deviations. We see that the typical variations are of a few percent. The trend for the US, China and France seems to be a slight increase in the ratio. We also notice that the ratio is negative on average for Denmark, meaning that the government typically has surpluses.}
    \label{fig:primarydef_to_GDP}
\end{figure}

This leads to the identity

\begin{equation}
    \gamma_{\tau} - 1 = \frac{\left(T_{\tau+1} + U_{\tau+1} + I(\tau+1) + X_{\tau+1} - M_{\tau+1}\right) - C_{\tau}}{C_{\tau}},
\end{equation}

which means that $\gamma_{\tau}$ encapsulates relative growth due to non-governmental activities, particularly if taxation is an increasing function of economic fluxes. With such definitions, we obtain

\begin{equation}
\begin{split}
    C_{\tau + 1} &= \gamma_{\tau} C_{\tau} + \left(G_{\tau+1} - T_{\tau+1}\right)
    \\ D_{\tau+1} &= \rho D_{\tau} + \left(G_{\tau+1} - T_{\tau+1}\right),
\end{split}
\end{equation}

which we can identify with the previous model a priori, under the condition that we define a leverage

\begin{equation}
    L_{\tau} = 1 + \frac{G_{\tau+1}-T_{\tau+1}}{C_{\tau}},
\end{equation}

which thus corresponds to the comparison between the primary deficit of year $\tau+1$ to the GDP of year $\tau$. Indeed, we can introduce the deficit $d_{\tau+1}=G_{\tau+1}-T_{\tau+1}$, and we get $L_{\tau} = 1 + d_{\tau+1}/C_{\tau}$. To assess the validity of our hypothesis that this leverage is fixed for a given country, we show this value from data for the US, China, France, and Denmark in Fig.\ref{fig:primarydef_to_GDP}. We observe a small variation (of a few percent) over the period of study, indeed.

\subsection{Relation to the Keen model}

The Keen model~\cite{keen_finance_1995,keen_predicting_2013, keen_emergent_2020, giraud_macrodynamics_2023} is a model inspired from Minsky, that describes the dynamics of macroeconomic variables. In that model, the equations for $C_{\tau}$ and $D_{\tau}$, take the following form in terms of the total growth rate $g_{\tau}$, the total interest rate $r_{\tau}$, the investment to output ratio $\kappa_{\tau}$, the wage to labor ratio $w_{\tau}$, and the output to labor ratio $a_{\tau}$:

\begin{equation}
\begin{split}
    C_{\tau+1}-C_{\tau} & = g_{\tau}C_{\tau}
    \\ D_{\tau+1}-D_{\tau} & = r_{\tau} D_{\tau} + \left(\kappa_{\tau}+\frac{w_{\tau}}{a_{\tau}}-1\right)C_{\tau}.
\end{split}
\end{equation}
This leads to the following mapping to our model
\begin{equation}
\begin{split}
    \gamma_{\tau} & \sim g_{\tau}-\left(\kappa_{\tau}+\frac{w_{\tau}}{a_{\tau}}\right)
    \\ \rho_{\tau} & \sim 1 + r_{\tau}
    \\ L_{\tau} & \sim \left(\kappa_{\tau}+\frac{w_{\tau}}{a_{\tau}}\right).
\end{split}
\end{equation}

In practice $w_{\tau}/a_{\tau}$ is the wage to output ratio, and $\kappa_{\tau}$ represents the share of output dedicated to investments. Following Fig.\ref{fig:primarydef_to_GDP}, this suggests that the yearly variation of $\kappa_{\tau}+w_{\tau}/a_{\tau}$ is weak. The stochasticity of our model is thus encapsulated in $g_{\tau}$ (which is typically deterministic in the Keen model \cite{keen_finance_1995, giraud_macrodynamics_2023}). This also gives insights on the variables, $L_{\tau}$ accounts for investments and labour.

\section{Results}

\subsection{Lock-in of carbon emissions}

Assuming that an external bank lends money each year at the cost of increasing the debt, we derive the evolution of a country's economy in terms of its GDP $C_{\tau}$. In practice, the capital will be used to invest in energetically costly activities. This means that we can associate to $C_{\tau}$ a function $e(C_{\tau})$ which represents the energetic cost of the GDP $C_{\tau}$. This should be an increasing function of the capital, and thus its average should increase with $L$ and $C_{\text{ini}}$. There exists a relationship between energy and economical value in dollars for different sectors, explored with Leontief's input output analysis in~\cite{costanza_embodied_1980}. If we also assume that emission, or dissipation $\epsilon$ is an increasing function of energy consumption, we ultimately have that $\epsilon$ is an increasing function of the capital $C_{\tau}$ (which is observed in practice~\cite{tucker_carbon_1995}). A second order relationship~\cite{tucker_carbon_1995} $\epsilon(C_{\tau}) = \alpha_0 + \alpha_1 C_{\tau} + \alpha_2 C_{\tau}$ can be used to model emissions encapsulate that emissions tend to increase faster as $GDP$ increases. The parameters $\alpha_{0,1,2}$ may depend on time, for instance if more renewable energies are used over time. We rather use the carbon intensity of an economy $I$~\cite{friedlingstein_persistent_2014}, representing the carbon cost of $1 USD$ (which depends on a country's activities, energy mix, .... This is given by Kaya's relationship~\cite{friedlingstein_persistent_2014, tavakoli_journey_2018,raupach_global_2007, abbasi_carbon_2022}
-
\begin{equation}
    \epsilon (\tau, C_{\tau}) = C_{\tau} \times I(\tau).
\end{equation}

The cumulative emissions of a country depend on the history of its economy, we refer to this effect as the \textit{lock-in effect}, and measure it using the \textit{path-dependent intensity}

\begin{equation}
    \mathfrak{I}_{\tau}\left(\{C_{\tau}\}\right) = \frac{\sum_{\tau'=0}^{\tau} I(\tau')C_{\tau'}}{\sum_{\tau'=0}^{\tau}C_{\tau'}},
\end{equation}

\subsection{Energetic cost of debt}

On Table.1 of the main text we report the different values extracted from data, for different countries, averaged over the period $1998-2022$. To obtain $L, \rho, W$, we use the following equations and data for $GDP$, debt, and deficit

\begin{equation}
\begin{split}
    L_{\tau} &= 1 + \frac{d_{\tau+1}}{C_{\tau}}
    \\\rho_{\tau} &= \frac{D_{\tau+1} - d_{\tau+1}}{D_{\tau}}
    \\\gamma_{\tau} &= \frac{C_{\tau+1} - d_{\tau+1}}{C_{\tau}},
\end{split}
\end{equation}

where $d_{\tau+1}$ is the deficit of the state during year $\tau$ for year $\tau+1$, and the fact that $W=\avg{\ln(\gamma_{\tau})}$ (we use the same definition as in model A).

% \begin{table}[h!]
%     \centering
%     \begin{tabular}{|c|c|c|c|c|}
%          \hline
%          & China & US & Denmark & France 
%          \\ \hline $\avg{L_{\tau}}$ & $1.025 \pm 0.023$ & $1.051 \pm 0.040$ & $0.997 \pm 0.029$ & $1.040 \pm 0.018$
         
%          \\ \hline $\avg{\rho_{\tau}}$ & $1.125 \pm 0.023$ & $1.010 \pm 0.051$ & $1.014 \pm 0.155$ & $1.002 \pm 0.091$ 
         
%          \\ \hline $W$ & $0.088 \pm 0.086$ & $-0.018 \pm 0.059$ & $0.032 \pm 0.089$ & $-0.020 \pm 0.090$ 
%          \\ \hline
%     \end{tabular}
%     \caption{Values of the different parameters.}
%     \label{tab:measur_parameters_avg}
% \end{table}

Considering a period $t$ we can write the cumulated emissions $\mathcal{E}(t)$ and resource consumptions $\mathcal{R}(t)$ as 
\begin{equation}
\begin{split}
    \mathcal{E}(t) &= \sum_{\tau \leq t} \epsilon(C_{\tau},\tau)
    \\ \mathcal{R}(t) &= \sum_{\tau \leq t}r(C_{\tau},\tau),
\end{split}
\end{equation}
where $\epsilon(C,\tau)$ are the emissions of investing $C$ at time $\tau$, and $r(C,\tau)$ is the total use in resource when investing $C$ at time $\tau$. For the US, we obtain Fig.\ref{fig:supp_GDP_debt} with the coefficients of determination $R^2 = 0.82$ for GDP and $R^2 = 0.79$ for debt. To ensure the model is working, we also use it to model GDP and debt for other countries and estimate their associated emissions. For France, we get the coefficients of determination $R^2 = 0.53$ for GDP and $R^2 = 0.86$ for debt. For China, we get the coefficients of determination $R^2 = 0.91$ for GDP and $R^2 = 0.97$ for debt. For Denmark, we obtain coefficients of determination $R^2 = 0.62$ for GDP and a negative $R^2$ for debt, indicating that the model fails to capture the strong variations and changes in sign of leverage. We observe a greater discrepancy between predictions and data for debt in Denmark. This is because Denmark usually runs a surplus and partially repays the debt. Therefore, the predicted debt exceeds the actual debt, which usually decreases. We report the coefficients of determination and rescaled mean absolute errors in Table \ref{tab:R2MAE_GDP_debt}.
Note that in this table, we used the notations:
\begin{equation}
    \frac{MAE_X}{\avg{X}} = \frac{\sum_{n=1}^{N_{m}}|X_n-X_n^{\text{simu}}|}{\sum_{n=1}^{N_{m}}X_n},
\end{equation}
with $N_m$ the number of measurements and 
\begin{equation}
    R^2_X = 1 - \frac{\sum_{n=1}^{N_{m}}(X_n-X_n^{\text{simu}})^2}{\sum_{n=1}^{N_{m}}(X_n-\avg{X})^2}.
\end{equation}

\begin{table}[h!]
    \centering
    \begin{tabular}{|c|c|c|c|c|}
         \hline
         & China & US & Denmark & France 
         \\ \hline $MAE_{GDP}/\avg{{GDP}}$ & $0.155$ & $0.081$ & $0.116$ & $0.119$
         
         \\ \hline $MAE_{\text{Debt}}/\avg{\text{Debt}}$ & $0.141$ & $0.203$ & $0.172$ & $0.119$
         \\ \hline $R^2_{GDP}$ & $0.91$ & $0.82$ & $0.62$ & $0.53$
         
         \\ \hline $R^2_{\text{Debt}}$ & $0.97$ & $0.79$ & $-1.22$ & $0.86$
         \\ \hline
    \end{tabular}
    \caption{Rescaled Mean Absolute Error $MAE$ and coefficient of determination $R^2$ between data and theory for GDP and debt, for different countries. 
    \label{tab:R2MAE_GDP_debt}}
\end{table}

\begin{figure}
    \centering
    \includegraphics[width=\textwidth]{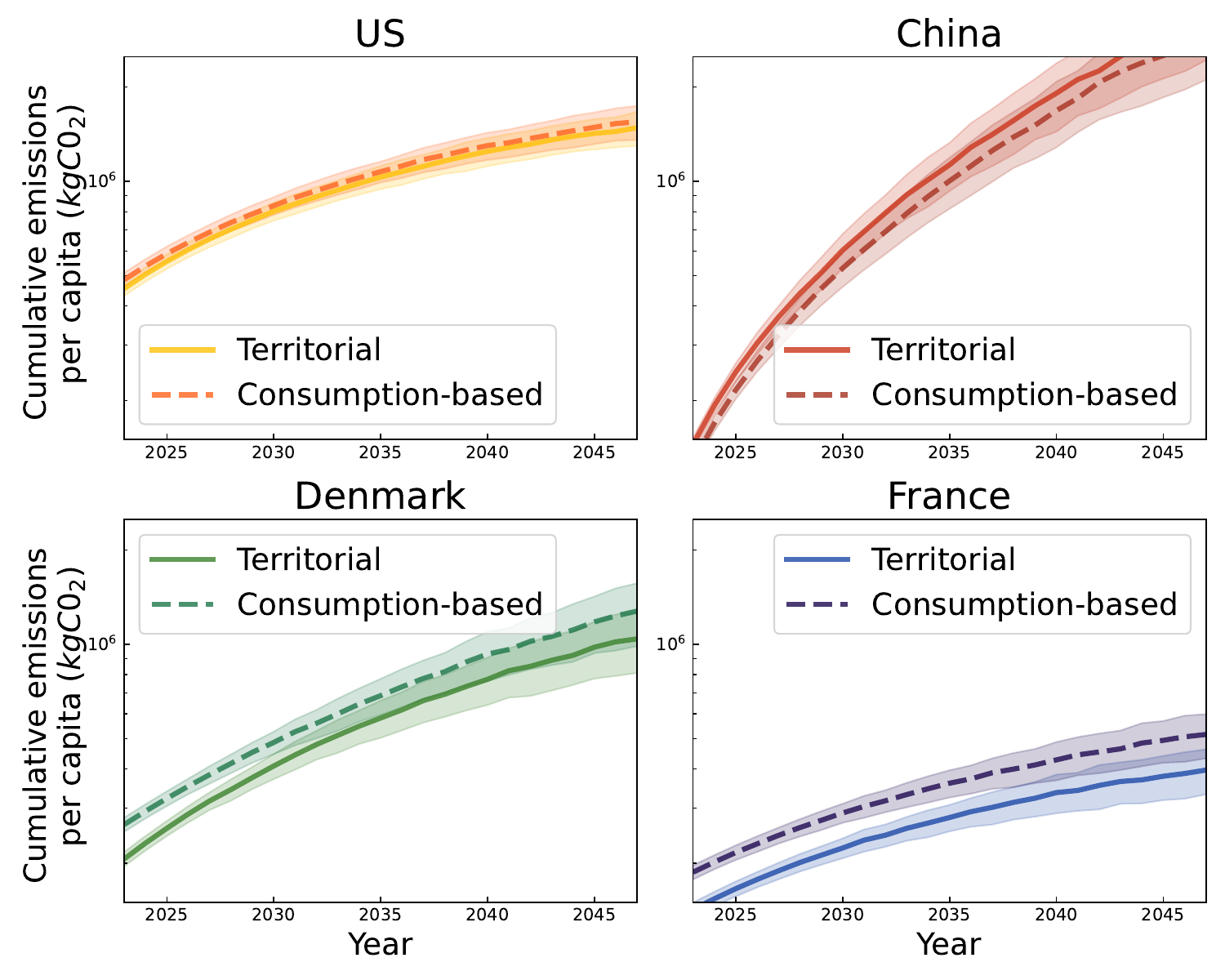}
    \caption{Comparing territorial emissions and consumption-based emissions using our model for predictions. For France and Denmark, we observe that territorial emissions are lower than consumption-based. Predictions are performed by running the model with the distributions of values for leverage $L$, $W$, $\rho$, and carbon intensity over the period 1998-2022. Carbon intensities are assumed to be linear in time, with a constant decarbonation rate corresponding to that of the period 1998-2022 (which is different for territorial emissions and consumption-based emissions).}
    \label{fig:territVSconsbas}
\end{figure}

\begin{figure}
    \centering
    \includegraphics[width=\linewidth]{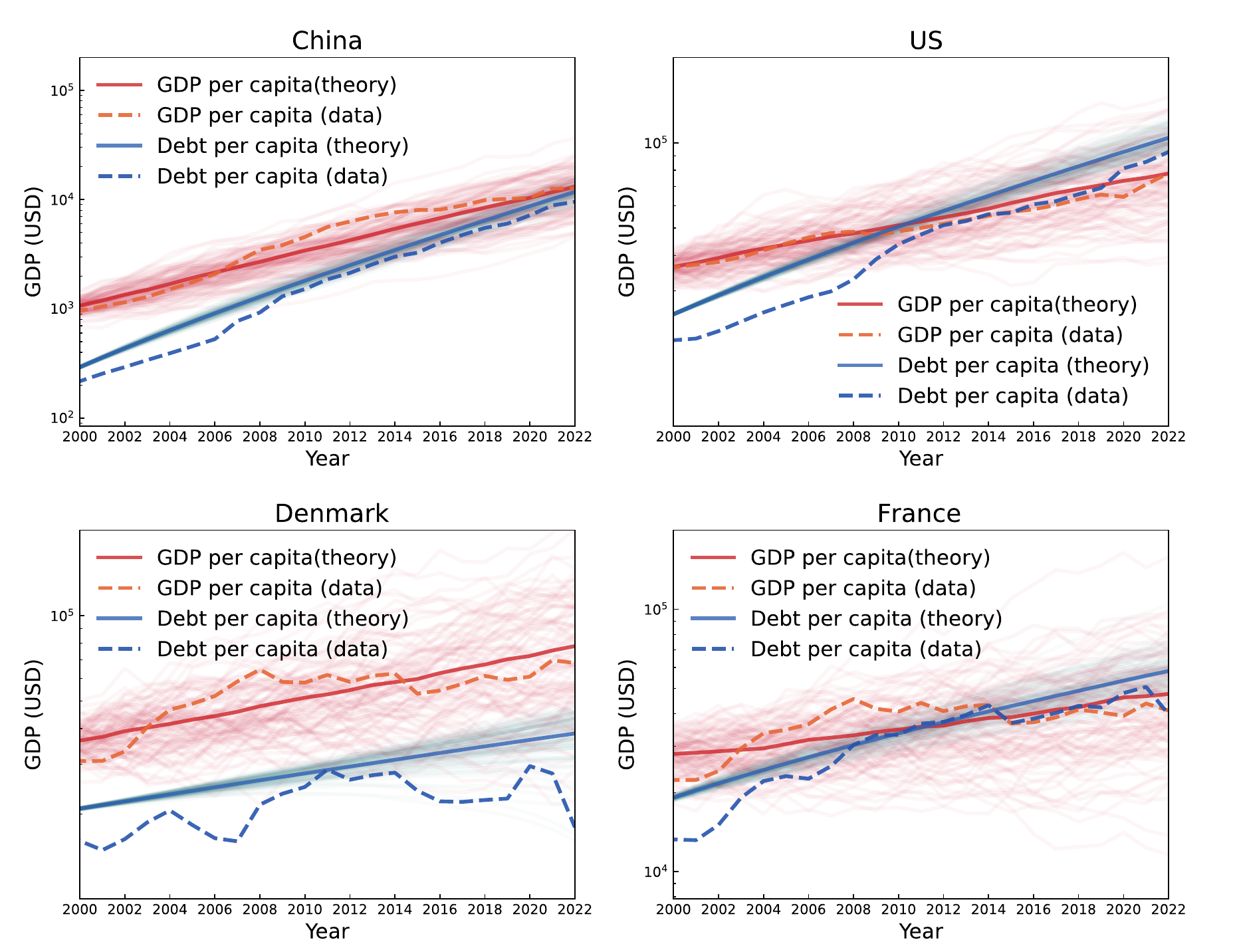}
        \caption{Comparison to data for $GDP$ per capita, and debt per capita between 2000 and 2020 for all countries. This is not a fit, but only using values for the deficit, GDP and debt to compute the parameters of the model.}
        \label{fig:supp_GDP_debt}
\end{figure}

It is important to understand the difference between territorial and consumption-based emissions. Territorial emissions represent carbon emissions produced on the territory of a given country, while consumption-based counts emissions due to activities directly used by a given country.  Thus, some countries may display a comparatively low carbon intensity by only accounting for territorial emissions, like France or Denmark. However, when accounting for exported emissions, their cumulative emissions increase. This is illustrated with predictions using our model in Fig.\ref{fig:territVSconsbas}.

\begin{figure}
    \centering
    \includegraphics[width=\linewidth]{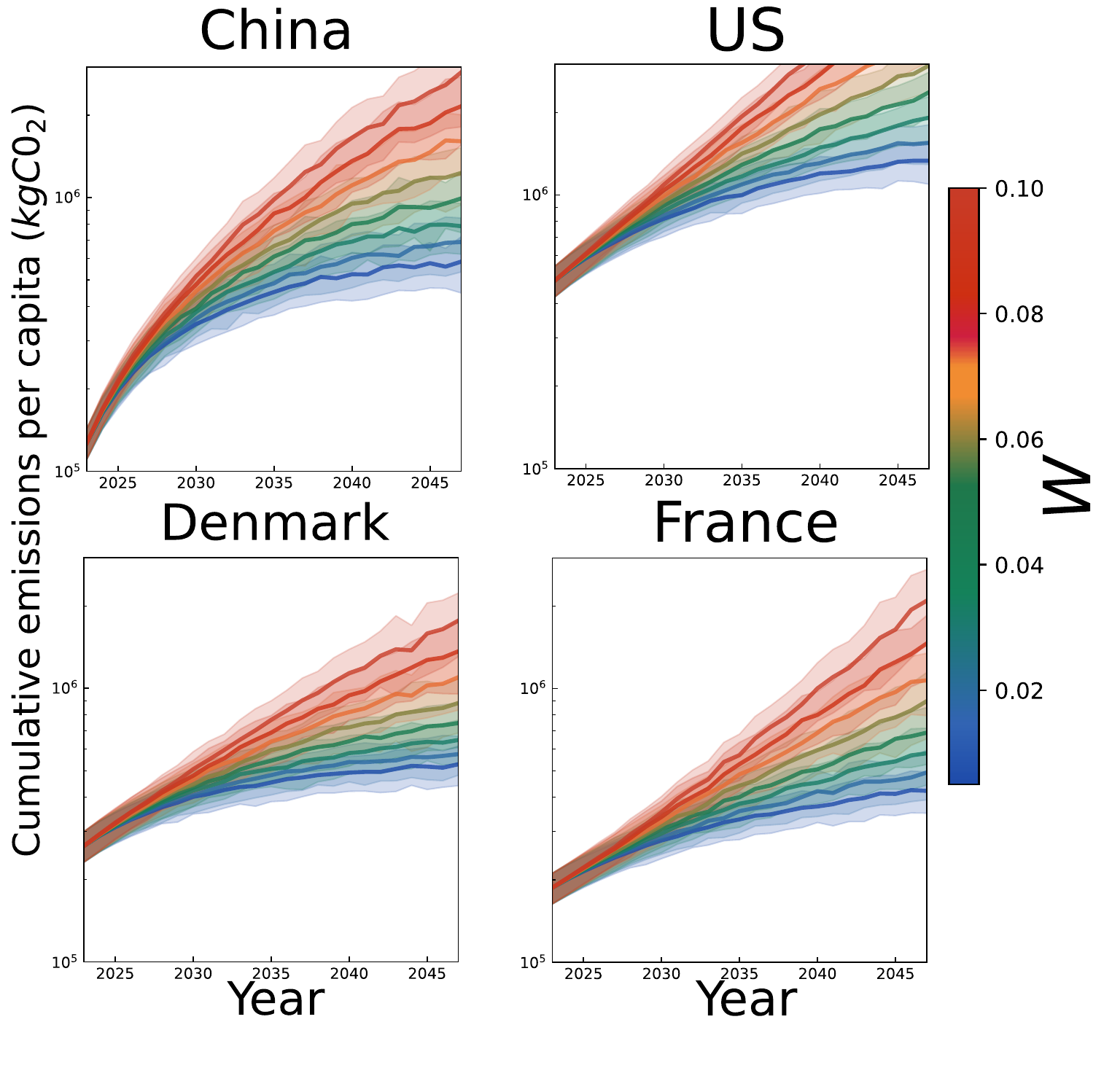}
    \caption{Predictions for cumulative emissions depending on the long-term growth rate $W$ in different countries. We used our stochastic credit-risk model and assume a linearly decreasing carbon intensity (with different parameters for each country). We assume that the leverage remains the same in each country.}
    \label{fig:pred_W}
\end{figure}

In Fig.\ref{fig:pred_W}, we show the predictions of the model for various intrinsic growth rates, which carries a similar message as Fig.3 of the main text.

\subsection{Comparing path dependent intensities to yearly intensities}

If we want to compare $\mathfrak{I}_{\tau}\left(\{C_{\tau}\}\right)$ to $I(\tau)$ in general, we get

\begin{equation}
\begin{split}
    \mathfrak{I}_{\tau}\left(\{C_{\tau}\}\right) - I(\tau) = \frac{\sum_{\tau'=0}^{\tau}\left(I(\tau')-I(\tau)\right)C_{\tau'}}{\sum_{\tau'=0}^{\tau}C_{\tau'}},
\end{split}
\end{equation}

from which we can easily see that if $\tau \to I(\tau)$ is a monotonically decreasing function, $\mathfrak{I}_{\tau}\left(\{C_{\tau}\}\right)$ will always be larger than $I(\tau)$. $I(\tau)$ is a decreasing function in many countries due to changes in energy use, often referred to as innovation. We can then compute the evolution of the quantity, 
\begin{equation}
\begin{split}
    \mathfrak{I}_{\tau+1}\left(\{C_{\tau}\}\right)-\mathfrak{I}_{\tau}\left(\{C_{\tau}\}\right) &= \frac{1}{S_{\tau}S_{\tau+1}}\left(\sum_{\tau'=0}^{\tau+1}I(\tau')S_{\tau}C_{\tau'} - \sum_{\tau'=0}^{\tau}I(\tau')S_{\tau+1}C_{\tau'}\right)
    \\ &= \frac{1}{S_{\tau}S_{\tau+1}}\left(I(\tau+1)C_{\tau+1}S_{\tau}+\sum_{\tau'=0}^{\tau}I(\tau')C_{\tau'}\left(S_{\tau}-S_{\tau+1}\right)\right)
    \\&= \frac{C_{\tau+1}}{S_{\tau+1}}\left(I(\tau+1)-\mathfrak{I}_{\tau}\left(\{C_{\tau}\}\right)\right),
\end{split}
\end{equation}
where we have introduced the notation $S_{\tau}=\sum_{\tau'=0}^{\tau}C_{\tau'}$.

The last equation leads to the conclusion that the path-dependent intensity is decreasing with time if and only if $I(\tau+1)<\mathfrak{I}_{\tau}\left(\{C_{\tau}\}\right)$ that is, if and only if

\begin{equation}
    \sum_{\tau'=0}^{\tau}\left(I(\tau')-I(\tau+1)\right)C_{\tau'}>0.
\end{equation}

This means that, as long as the yearly carbon intensity is a decreasing function, the path-dependent intensity is as well. For a non-monotonic intensity, the years with the highest $C_{\tau}$ will prevail in the sum. From this we can also deduce the variations of $\Delta_{\tau} := \mathfrak{I}_{\tau}\left(\{C_{\tau}\}\right) - I(\tau)$,

\begin{equation}
\begin{split}
    \Delta_{\tau+1}-\Delta_{\tau} &= \frac{C_{\tau+1}}{S_{\tau+1}}\left(I(\tau+1)-\mathfrak{I}_{\tau}\left(\{C_{\tau}\}\right)\right) - (I(\tau+1)-I(\tau))
    \\ &= \frac{S_{\tau}}{S_{\tau+1}}\left(I(\tau) - I(\tau+1)\right) -\frac{C_{\tau+1}}{S_{\tau+1}}\Delta_{\tau},
\end{split}
\end{equation}

which also means that for $I(\tau)$ strictly decreasing, $\Delta_{\tau}$ is initially increasing driven by the first term $S_{\tau}\left(I(\tau) - I(\tau+1)\right)/S_{\tau+1}$. Then it starts decreasing for $S_{\tau}\left(I(\tau) - I(\tau+1)\right)< C_{\tau+1}\Delta_{\tau}$. In the deterministic limit, we approximate $C_{\tau}\sim e^{W\tau}$ at large $\tau$, where $W$ is the growth rate. Thus, we obtain that this condition is

\begin{equation}
    \Delta_{\tau} > \left(I(\tau) - I(\tau+1)\right)\frac{1-e^{-W(\tau+1)}}{e^{W} - 1}.
\end{equation}

For a decreasing yearly intensity, $\Delta_{\tau}$ first increases until this condition is reached, then evolves depending on the sign of $\left(I(\tau) - I(\tau+1)\right)$. Then at large times, we get 

\begin{equation}
    \Delta_{\tau+1} \sim \left(I(\tau) - I(\tau+1) + \Delta_{\tau}\right)e^{-W}.
\end{equation}

From this, we deduce that at large times

\begin{equation}
    \Delta_{\tau} \sim -\frac{1}{e^{W}-1}\frac{dI}{d\tau},
\end{equation}

meaning that $\mathfrak{I}_{\tau}\left(\{C_{\tau}\}\right) \sim I(\tau)-(dI/d\tau)/(e^{W}-1)$. In particular if we assume a power law dependency for the carbon intensity of the form $I(\tau) = I_0 - \eta \tau^n$, the difference between $\mathfrak{I}_{\tau}\left(\{C_{\tau}\}\right)$ and $I(\tau)$ goes as 

\begin{equation}
    \Delta_{\tau} \sim \frac{\eta n \tau^{n-1}}{e^{W}-1},
\end{equation}
therefore increasing to $+\infty$ for $n>1$ and decreasing to $0$ for $n<1$ as $\tau \to \infty$. In practice, it means that for fast decarbonization, early intense years matter more in the path-dependent intensity.

\paragraph{Illustration with linearly decreasing intensities}

If we use a linear relationship $I(\tau)=I_0 - \eta (\tau-\tau_0)$ (which seems to be observed in most cases~\cite{friedlingstein_persistent_2014}), we get that

\begin{equation}
    \mathfrak{I}_{\tau}\left(\{C_{\tau}\}\right) = I_0 - \eta \frac{\sum_{\tau'=0}^{\tau}\tau'C_{\tau'}}{\sum_{\tau'=0}^{\tau}C_{\tau'}},
\end{equation}

\begin{figure}
    \centering
    \includegraphics[width=\linewidth]{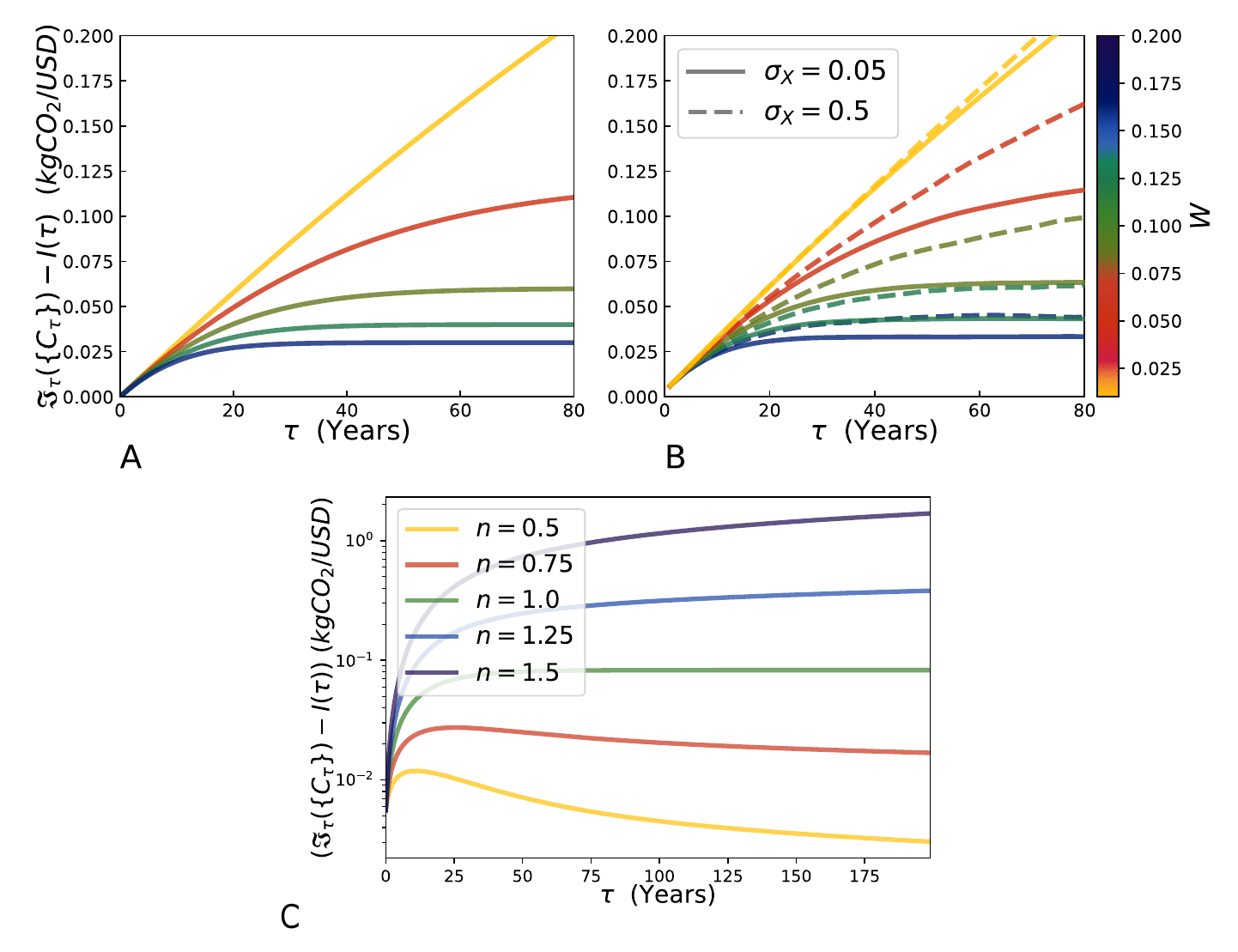}
    \caption{\textbf{A} Difference between the path dependent intensity $\mathfrak{I}_{\tau}\left(\{C_{\tau}\}\right)$ and the yearly intensity $I(\tau)$ with a deterministic evolution of $C_{\tau}$. The path-dependent intensity is systematically larger than the yearly intensity for $W>0$ and increases with time. It eventually reaches a strictly positive limit. \textbf{B} Effect of fluctuations in the economy on the difference between the path-dependent intensity $\mathfrak{I}_{\tau}\left(\{C_{\tau}\}\right)$ and the yearly intensity $I(\tau)$. \textbf{C} $\mathfrak{I}_{\tau}\left(\{C_{\tau}\}\right) - I(\tau)$ follows a power law $n\eta \tau^{n-1}/(e^{W}-1)$ after a first transient phase, depending on the decarbonization rate.}
    \label{fig:determ_intensities}
\end{figure}

In addition, if we forget about stochasticity and assume exponential growth $C_{\tau} = LC_{\text{ini}} e^{W\tau}$, we get

\begin{equation}
    \mathfrak{I}_{\tau}\left(\{C_{\tau}\}\right) = I_0 - \frac{\eta e^{W}}{e^{W} - 1}\left(\frac{e^{W\tau}(\tau+1)(e^{W}-1)}{e^{W(\tau+1)}-1} - 1\right),
\end{equation}

and finally

\begin{equation}
    \mathfrak{I}_{\tau}\left(\{C_{\tau}\}\right) - I(\tau) = \eta\left( \tau + \frac{e^{W}((\tau+1)e^{W\tau}-\tau e^{W(\tau+1)}-1))}{(e^{W}-1)(e^{W(\tau+1)}-1)} \right),
\end{equation}

which is positive and increasing towards $1/(e^{W}-1)$ provided that $W>0$ as can be seen on Fig.\ref{fig:determ_intensities}. This means that path-dependent intensities are always larger than yearly intensities whenever GDP is growing over time.

\paragraph{With stochasticity}

Let us now add stochasticity and introduce the stochastic variable $V_\tau=\gamma_\tau + L-1$ and $\pi_t = \prod_{1\leq \tau'\leq\tau} V_{\tau'}$ so that the capital is $C_{\tau}=LC_{\text{ini}}\pi(\tau)$. Let us also introduce the random variable $X_{\tau'} = \ln(V_{\tau'})$, so that $C_{\tau}=LC_{\text{ini}}\prod_{1\leq\tau'\leq\tau}e^{X_{\tau'}}$. In the limit where $X_{\tau'}$ is drawn according to a normal distribution with mean $W$ and standard deviation $\sigma_X$, we write $X_{\tau'}=W+\sigma_X \xi_{\tau'}$ ($\xi_{\tau'}$ is normally distributed with mean $0$ and standard deviation $1$). We can thus write

\begin{equation}
\begin{split}
    \mathfrak{I}_{\tau}\left(\{C_{\tau}\}\right) - I(\tau) &= \frac{\sum_{\tau'=0}^{\tau}\left(I(\tau')-I(\tau)\right)e^{W\tau'}\prod_{1\leq\tau''\leq\tau'}e^{\sigma_X\xi_{\tau''}}}{\sum_{\tau'=0}^{\tau}e^{W\tau'}\prod_{1\leq\tau''\leq\tau'}e^{\sigma_X\xi_{\tau''}}}
    \\&=\frac{\sum_{\tau'=0}^{\tau}\left(I(\tau')-I(\tau)\right)\exp\left(W\tau'+\sigma_X\sum_{1\leq\tau''\leq\tau'}\xi_{\tau''}\right)}{\sum_{\tau'=0}^{\tau}\exp\left(W\tau'+\sigma_X\sum_{1\leq\tau''\leq\tau'}\xi_{\tau''}\right)}.
\end{split}
\end{equation}

Since $\xi(\tau)$ is normally distributed, we get that $\avg{\exp\left(W\tau'+\sigma_X\sum_{1\leq\tau''\leq\tau'}\xi_{\tau''}\right)}=\exp\left((W+\sigma_X^2/2)\tau'\right)$, and 

\begin{equation}
    \mathbb{V} \left(\exp\left(W\tau'+\sigma_X\sum_{1\leq\tau''\leq\tau'}\xi_{\tau''}\right) \right) = \exp\left((2W+\sigma_X^2)\tau'\right)\left(\exp\left(\sigma_X^2\tau'\right)-1\right).
\end{equation}

Therefore, this variable has wide fluctuations. For small $\sigma_X$, we recover the previous deterministic result. We show the dependency of this on the fluctuations $\sigma_X$ in Fig.\ref{fig:determ_intensities}B. We observe that the difference between the path-dependent intensity and the yearly intensity increases with the fluctuations $\sigma_X$. This results from the fact that if a period of high economic growth with high carbon intensities is followed by a period of recessions with lower carbon intensities (years with a stochastically low growth rate), the weight of carbon-emissive years will be larger. Therefore, the weights of years with large carbon intensities will be larger and the difference between $\mathfrak{I}_{\tau}\left(\{C_{\tau}\}\right)$ and $I(\tau)$ will increase. In addition, large fluctuations may lead to years with a stochastically significant growth rate and low carbon intensities, typically reducing the difference between $\mathfrak{I}_{\tau}\left(\{C_{\tau}\}\right)$ and $I(\tau)$. However, as shown in the deterministic case, the difference typically scales as $1/W$, so that decreasing the growth rate stochastically has a larger impact on the difference between $\mathfrak{I}_{\tau}\left(\{C_{\tau}\}\right)$ and $I(\tau)$ than increasing it stochastically. Therefore, recessions lead to a enhanced difference between $\mathfrak{I}_{\tau}\left(\{C_{\tau}\}\right)$ and $I(\tau)$, which average rises when fluctuations are stronger.

To better understand this effect, we can study what happens for a first phase with a high growth rate $\gamma_h$, followed by a phase of recession with a lower growth rate $\gamma_l$. In this case, for a phase of fast growth lasting for $\tau_h$ and a phase of recession lasting for $\tau - \tau_h$, we get

\begin{equation}
\begin{split}
    \mathfrak{I}_{\tau}\left(\{C_{\tau}\}\right) &- I(\tau) = \frac{(e^{\gamma_h}-1)(e^{\gamma_l}-1)}{(e^{\gamma_l}-1)(e^{\gamma_l(\tau_h+1)}-1)+(e^{\gamma_h}-1)(e^{\gamma_h(\tau+1)}-e^{\gamma_h(\tau_h+1)})}\times
    \\ &\left(\frac{\left((e^{\gamma_h}-1)((\tau-\tau_l)e^{\gamma_h(\tau_h+1)} - \tau)-e^{\gamma_h}(1+e^{\gamma_h\tau_h})\right)}{(e^{\gamma_h}-1)^2}+\frac{\left((e^{\gamma_l}-1)(\tau_l-\tau)e^{\gamma_l(\tau_h+1)}-e^{\gamma_l(\tau+1)}+e^{\gamma_l(\tau_h+1)}\right)}{(e^{\gamma_l}-1)^2}\right),
\end{split}
\end{equation}

So that for large $\tau$, it scales as the duration of the recession divided by $e^{\gamma_l}-1$

\begin{equation}
    \mathfrak{I}_{\tau}\left(\{C_{\tau}\}\right) - I(\tau) \sim \frac{\tau-\tau_h}{e^{\gamma_l}-1}
\end{equation}

Therefore, only the growth rate and duration of the recession phase matter to explain the difference in trends between $\mathfrak{I}_{\tau}\left(\{C_{\tau}\}\right)$ and $I(\tau)$ in this limit.

\subsection{Study of the solvency probability}

\paragraph{Strategy A}
 We can eventually relate this to the probability that a biased random walk is below a threshold at payback time $t_p$~\cite{hod_survival_2020, majumdar_universal_2010} (the jumps do not have zero mean, and the random walk is thus biased). We emphasize that bankruptcy does not correspond to the existence of a time where the debt is larger than the gross capital (or first passage time), but to the fact that at the time of payback $t_p$, the debt is larger than the gross capital.

To evaluate it, let us start by introducing the stochastic variable $X_{\tau'} = \ln(\gamma_{\tau'} / \rho)$, which has mean $m=W-\ln(\rho) $ and variance $\sigma_X^2$. Using the central limit theorem for large enough $\tau$, the solvency probability is 
\begin{equation}
\begin{split}
P_S^A(t_p)&=\mathbb{P}\left(\sum_{\tau'\leq\tau}X_{\tau'}\geq \ln\left(1-\frac{1}{L}\right)\right)
    \\ &=\frac{1}{2}\left(1+\text{erf}\left(\frac{t_p m -\ln\left(1-\frac{1}{L}\right)}{\sqrt{2t_p}\sigma_X}\right) \right).
\end{split}
\end{equation}

\begin{figure}
    \centering
    \includegraphics[width=\linewidth]{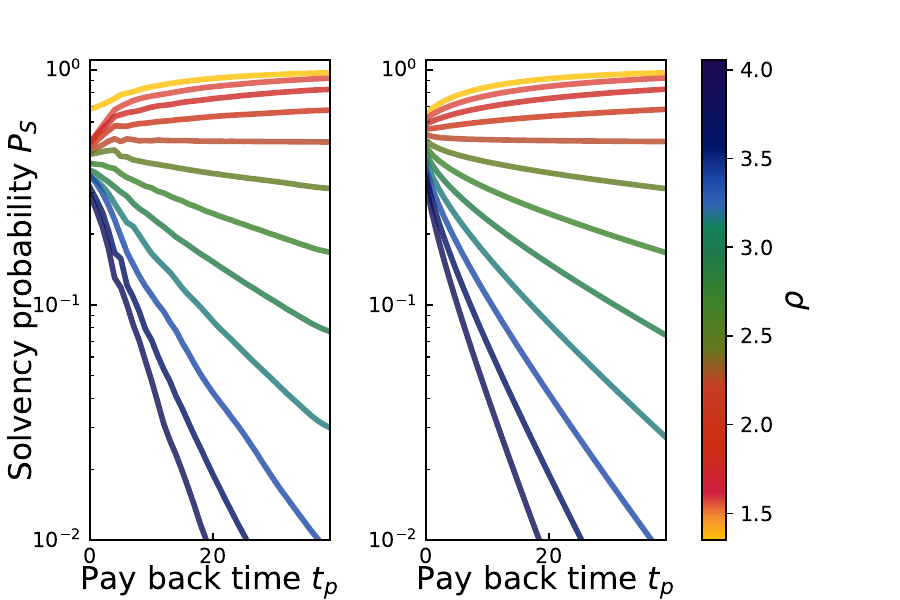}
    \caption{Solvency probability with Gaussian jumps, but the same average and standard deviation as in Kelly's case, in the limit of large leverage $L=10$ with strategy A.}
    \label{fig:gaussian_survprob_A}
\end{figure}

From this expression, we find that 
%if $m>0$ ( $W>\ln(\rho)$), it is a non-monotonic function, decreasing until $t_p = -\ln(1-1/L)/(W-\ln(\rho))$ (worst payback time) and then increasing towards $1$. Otherwise, it is a strictly decreasing function of time. Therefore, 
the solvency probability strongly depends on the sign of $m$. We plot this solvency probability as a function of the payback time $t_p$ in Fig.\ref{fig:gaussian_survprob_A} in the limit of large leverage. Using the expression of the short-term growth rate with strategy A given in Eq. \ref{eq:w_t_p}, we have that this is defined for all times if $E\geq\rho$. Otherwise, this is well defined for all times lower than $-\ln(1-1/L)/(\ln(\rho)-\ln(E))$. 
% In addition, if we define 

% \begin{equation}
%     g(t_p) = E^{t_p}-(1-\frac{1}{L})\rho^{t_p},
% \end{equation}

% which is the argument of the $\ln$ in the definition of the short-term growth rate, we have that

% \begin{equation}
%     w'(t_p) = \frac{1}{t_p^2}\left(t_p\frac{g'(t_p)}{g(t_p)}-\ln(g(t_p))\right).
% \end{equation}

% The latter is equal to $0$ if and only if $t_pg'(t_p)/g(t_p)=\ln(g(t_p))$. 
Regarding the solvency probability, we observe that it is a decreasing function of leverage (higher leverage is associated with higher risk). We thus identify four regimes based on the concavity inequality $\ln(E)\geq W$ :

\begin{itemize}
    \item If $m>0$ and $\ln(E)\geq\ln(\rho)$, $P_S^A(t_p\to\infty) \to 1$, faster as $\sigma_X$ decreases. The solvency probability is a non monotonic function of $t_p$, first decreasing until $t_p=-\ln(1-1/L)/(W-\ln(\rho))$, then increasing towards $1$. For large leverage, the first regime vanishes (when $-\ln(1-1/L)/(W-\ln(\rho))<1$), but it means that for small enough leverage, it may be safer to pay back as soon as possible, or to wait for long enough. The short-term growth rate is well defined at all times.
    \item If $m=0$ and $\ln(E)\geq\ln(\rho)$, $P_S^A(t_p\to\infty) = \frac{1}{2}$. In this case, the solvency probability is decreasing with $t_p$. The short-term growth rate is well defined at all times.
    \item If $m<0$ and $\ln(E)\geq\ln(\rho)$, $P_S^A(t_p\to\infty) \to 0$, faster as $\sigma_X$ decreases. The solvency probability is decreasing with $t_p$. The short-term growth rate is well defined at all times.
    \item If $m<0$ and $\ln(E)<\ln(\rho)$, $P_S^A(t_p\to\infty) \to 0$, faster as $\sigma_X$ decreases. The probability of solving decreases with $t_p$. The short-term growth rate is only defined for $t_p<-\ln(1-1/L)/(\ln(\rho)-\ln(E))$.
\end{itemize}

% Those results are summed up in Fig.\ref{fig:wtpPSVSL}.

% \begin{figure}
%     \includegraphics[width=\linewidth]{wtpPSVStp.pdf}
%     \caption{solvency probability and short-term growth rate for strategy A, different values of leverage and interest rates. For the solvency probability, there is a transition at $\ln(\rho)=W$. When $W<\ln(\rho)$, the solvency probability is strictly decreasing for all times; otherwise, it is a non-monotonic function of time, with a minimum at $t_p = -\ln(1-1/L)/(W-\ln(\rho))$. For the short-term growth rate, there is a transition at $\rho=E$. When the interest rate increases above $E$ and remains below $E^{L/(L-1)}$, the short-term growth rate starts decreasing with time after $t_p > \ln((1-1/L)\ln(\rho)/\ln(E))/\ln(E/\rho)$. The figure on the left corresponds to $\ln(\rho)<W$. The figure on the right corresponds to $W<\ln(\rho)$}
%     \label{fig:wtpPSVSL}
% \end{figure}

\paragraph{Strategy B}

Using strategy B, the solvency probability has a more complex expression, and we can not simply it directly with the central limit theorem. From Eq. \ref{eq:solvency_prob_B}, we have:

\begin{equation}
\begin{split}
    P_S^B(t_p) &= \mathbb{P}\left(\prod_{1\leq\tau'\leq t_p} \frac{1}{\rho}\left(\gamma_{\tau'} + (L-1)\right) -\sum_{0\leq t\leq t_p - 1}\frac{L-1}{\rho}\prod_{1\leq\tau'\leq t} \frac{1}{\rho}\left(\gamma_{\tau'} + (L-1)\right) \geq 1-\frac{1}{L}\right),
\end{split}
\end{equation}
which suggests to introduce the quantity $\Pi_t = \prod_{1\leq\tau'\leq t} \left(\gamma_{\tau'} + (L-1)\right)/\rho$, so that

\begin{equation}
\begin{split}
    P_S^B(t_p) &= \mathbb{P}\left(\Pi_{t_p} -\frac{L-1}{\rho}\sum_{0\leq t\leq t_p - 1} \Pi_t \geq 1-\frac{1}{L}\right),
\end{split}
\end{equation}
which is Eq. (14) of the main text with $X_{t_p}=\Pi_{t_p} -\frac{L-1}{\rho}\sum_{0\leq t\leq t_p - 1} \Pi_t$. 

Then, we have:
\begin{equation}
\begin{split}
    C_{t_p} &= LC_{\text{ini}} \rho^{t_p} \Pi_{t_p}
    \\ D_{t_p} &= LC_{\text{ini}} \rho^{t_p} \left(1-\frac{1}{L}+\frac{L-1}{\rho}\sum_{0\leq t \leq t_p-1}\Pi_{t}\right).
\end{split}
\end{equation}
Now we can write 
\begin{equation}
    \Pi(\tau) = \exp\left(\sum_{1\leq \tau'\leq\tau}\ln(V_{\tau'})\right) = \prod_{1\leq \tau'\leq\tau} V_{\tau'},
\end{equation}
and using the Central limit theorem, $\sum_{\tau'\leq\tau}\ln(V_{\tau'})$ follows a normal law $\mathcal{N}\left(\tau \mu, \sqrt{\tau}\sigma_X\right)$. Therefore $\Pi(\tau)$ is log-normally distributed, with mean $\exp\left(\tau(\mu+\sigma_X^2/2)\right)$ and variance $(\exp(\tau \sigma_X^2) - 1)\exp(\tau(2\mu + \sigma_X^2))$. Its probability distribution function is:

\begin{equation}
    \pi(\tau)(x) = \frac{1}{\sqrt{2\pi \tau}x\sigma_X}\exp\left(-\frac{(\ln(x)-\mu\tau)^2}{2\tau\sigma_X^2}\right).
\end{equation}

Now, we can rewrite $X_{t_p}$ in terms of the capital $C_{t_p}$ and debt $D_{t_p}$ at time $t_p$ as
\begin{equation}
\begin{split}
    X_{t_p} &=\frac{C_{t_p} - D_{t_p}}{LC_{\text{ini}} \rho^{t_p}} + 1 - \frac{1}{L}
    \\ &= \Pi_{t_p} -\frac{L-1}{\rho}\sum_{0\leq t\leq t_p - 1} \Pi_t
    \\ &= V_1 ... V_{t_p} - \frac{L-1}{\rho}\left(1 + V_1 + V_1 V_2 + ... + V_1 ... V_{t_p - 1}\right)
    \\ &= -\frac{L-1}{\rho} + V_1 \left( V_2 ... V_{t_p} - \frac{L-1}{\rho}\left(1 + V_2 + V_2 V_3 + ... + V_2 ... V_{t_p - 1}\right) \right)
    \\ &= -\frac{L-1}{\rho} + V_1 \left( - \frac{L-1}{\rho} + V_2 \left( V_3... V_{t_p} - \frac{L-1}{\rho}\left(1 + V_3 + V_3 V_4 + ... + V_3 ... V_{t_p - 1}\right)\right) \right)
    \\ &= -\frac{L-1}{\rho} + V_1\left(-\frac{L-1}{\rho} + V_2\left(-\frac{L-1}{\rho}+V_3\left(-\frac{L-1}{\rho}+...(-\frac{L-1}{\rho}+V_{t_p})\right)\right)\right),
\end{split}
\end{equation}

The sign of $X_{t_p}-(1-1/L)$ encapsulates solvency at time $t_p$. We can study the monotonicity of $X_{t_p}$, in particular we have that 

\begin{equation}
\begin{split}
    X_{t_p+1} - X_{t_p} &= (V_{tp+1} - 1)V_0...V_{t_p} -\frac{L-1}{\rho}V_0...V_{t_p}
    \\ &= V_0 V_1 ... V_{t_p}\left(V_{t_p+1}-\left(1+\frac{L-1}{\rho}\right)\right)
    \\ &= V_0 V_1 ... V_{t_p}\left(\frac{\gamma_{t_p+1}}{\rho}-1\right).
\end{split}
\end{equation}

Therefore, using the fact that the $V_{\tau}$ and $\gamma_{\tau}$ are independent, we get that on average the sign of $X_{t_p+1} - X_{t_p}$ is set by the sign of $\avg{\gamma}-\rho$.  We can thus draw two conclusions
\begin{itemize}
    \item If $\avg{\gamma}<\rho$ (which seems to be the case in most countries), we have that if the debt is larger than the capital at time $t_p$, then it is also the case at time $t_p +1$ on average. For a fully deterministic model, it would mean that the survival probability is a decreasing function of leverage. However, due to stochasticity, this result is no longer true when $\gamma$ has a broad distribution (the distribution is broader for higher $W$).

    \begin{equation}
    \begin{split}
        \left(\avg{X_{t_p+1}}<1-\frac{1}{L}\right)\subset\left(\avg{X_{t_p}}<1-\frac{1}{L}\right)
    \end{split}
    \end{equation}
    
    \item Otherwise if $\avg{\gamma}>\rho$, we have that if the debt is lower than the capital at time $t_p$ then it is also the case at time $t_p +1$ in average

    \begin{equation}
        \left(\avg{X_{t_p+1}}>1-\frac{1}{L}\right)\subset\left(\avg{X_{t_p}}>1-\frac{1}{L}\right)
    \end{equation}
\end{itemize}

In addition, the solvency probability is related to the cumulative distribution function of a sum of correlated log-normally distributed variables. If we assume that the jumps are normally distributed, $V_{\tau}$ are log-normally distributed and independent. We introduce the probability distribution function of a log-normal distribution

\begin{equation}
    \phi(x) = \frac{1}{\sqrt{2\pi}x\sigma_X}\exp\left(-\frac{(\ln(x)-\mu)^2}{2\sigma_X^2}\right).
\end{equation}

\begin{figure}
    \centering
    \includegraphics[width=\linewidth]{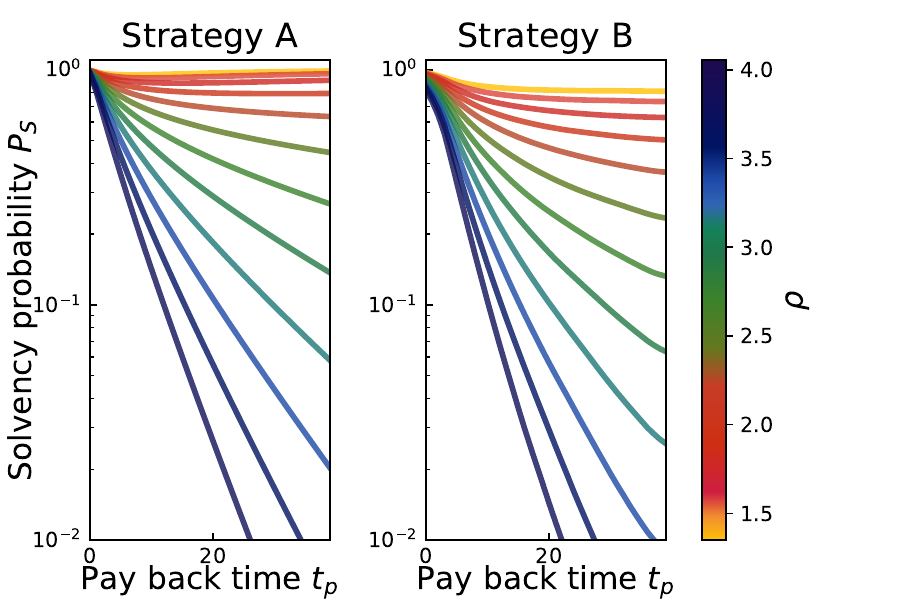}
    \caption{Comparing strategies with the same leverage ($L=1.05$), as expected, strategy A has a higher solvency probability. Yet it also leads to a lower capital in the end.}
    \label{fig:survprob_strategies}
\end{figure}

We plot it on Fig.\ref{fig:survprob_strategies}, where it reaches a plateau for small enough interest rates. We see that solvency is smaller with strategy B, a risk that has to be taken to sustain a larger growth rate. For $L \sim 1$ we understand why this solvency probability (strategy B) is smaller than for strategy A, because the investor in strategy B relies more on leverage than in strategy A.

\subsection{Optimal leverage and carbon emissions}

For strategy B, we get that the average growth rate is increasing with the leverage $\sim \avg{\log\left(\gamma_t+L-1\right)}$, while the survival probability has a non monotonous complex behavior as a function of leverage. Let us introduce an optimization function that interpolates between the growth rate and the survival probability:
\begin{equation}
    J(\beta, L, t_p) = \beta\avg{\log\left(\gamma_t+L-1\right)} + (1-\beta)P_S^B(t_p),
\end{equation}
where $\beta$ accounts for the relative weight of both contributions.

\begin{figure}
    \centering
    \includegraphics[width=\linewidth]{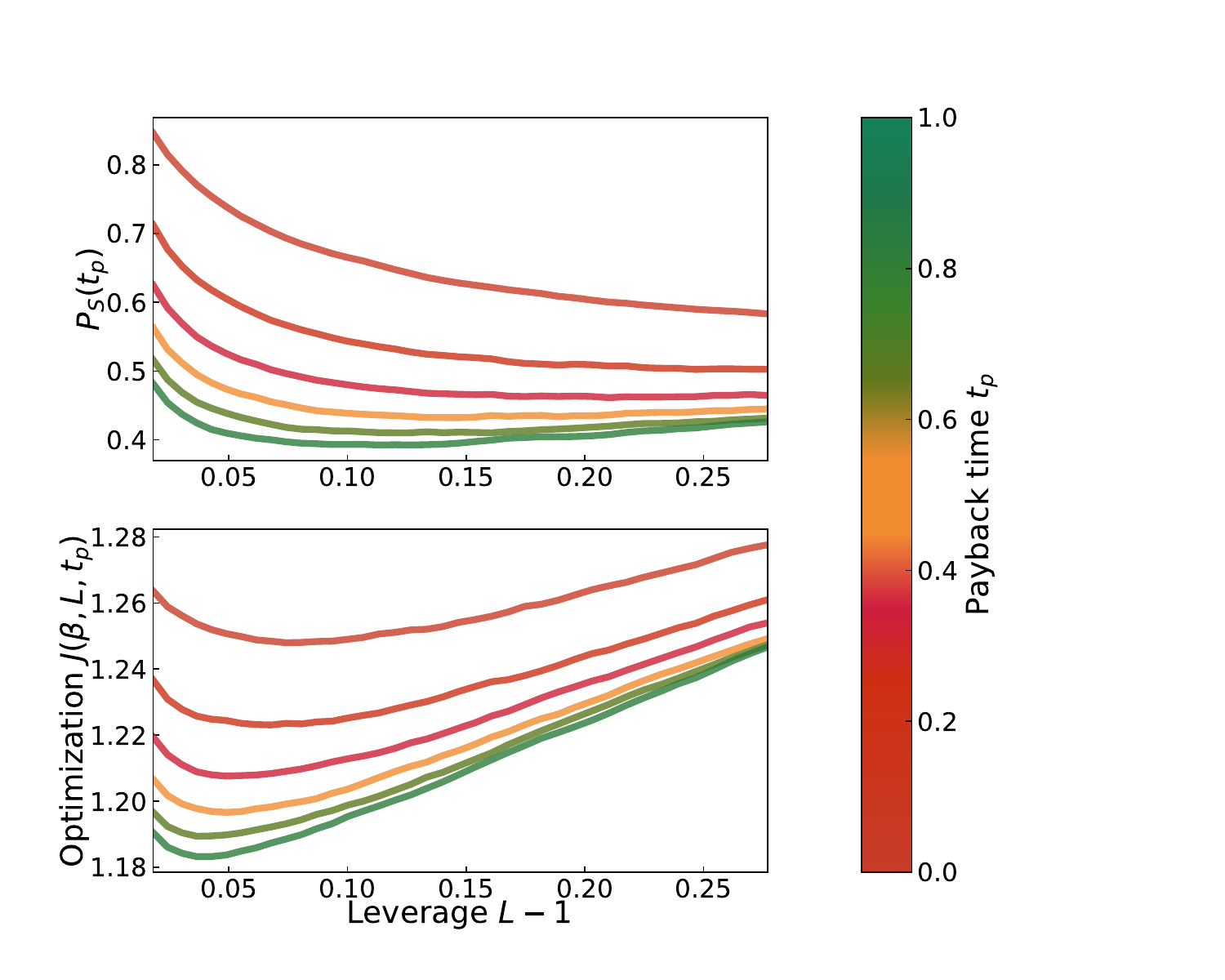}
    \caption{Optimization function for the leverage with $\beta=0.8, \rho=2, \avg{\log(\gamma)}=0.7$ and for different values of $t_p$. Decreasing $\beta$ leads to a stronger risk aversion, in which case the minimum of the optimization function is shifted to larger leverages. However the existence of a minimum does not depend on the value of $\beta$.}
    \label{fig:fig_opti_lev}
\end{figure}

Typically for the range of values corresponding to real data, we obtain the behavior of Fig.~\ref{fig:fig_opti_lev}. While the survival probability first decreases with leverage then plateaus or increases slowly, the growth rate keeps increasing with leverage. Therefore, the optimization function first decreases, reaches a minimum, and then increases again (the increase in growth rate compensates the plateau in survival probability). From this, we deduce that high leverages could be considered as optimal. However this optimization function does not account for carbon emissions resulting from a stronger growth rate. This suggests that the "optimal" strategy with credit in a carbon intensive economy leads to stronger cumulative emissions than without credit.

\section{Data}

All data used in this work are publicly available for the different countries under study. In particular, we get data for the US in~\cite{noauthor_us_2025, noauthor_historical_2025, noauthor_monthly_2025, noauthor_us_2025-1}. For France, we use~\cite{noauthor_france_2025, noauthor_france_2025-1, noauthor_france_2025-2, noauthor_france_2025-3}.  For Denmark, we use~\cite{noauthor_denmark_2025, noauthor_denmark_2025-1, noauthor_denmark_2025-2, noauthor_denmark_2025-3}.  For China, we use~\cite{noauthor_china_2025, noauthor_china_2025-1, noauthor_china_2025-2, noauthor_china_2025-3}.

%REFERENCES


\end{document}